\documentclass[a4paper,fleqn]{cas-dc}

\usepackage[authoryear]{natbib}

\usepackage{xargs} 
\usepackage[dvipsnames]{xcolor}  

\usepackage[colorinlistoftodos,prependcaption,textsize=tiny]{todonotes}
\newcommandx{\unsure}[2][1=]{\todo[linecolor=red,backgroundcolor=red!25,bordercolor=red,#1]{#2}}
\newcommandx{\change}[2][1=]{\todo[linecolor=blue,backgroundcolor=blue!25,bordercolor=blue,#1]{#2}}
\newcommandx{\info}[2][1=]{\todo[linecolor=OliveGreen,backgroundcolor=OliveGreen!25,bordercolor=OliveGreen,#1]{#2}}
\newcommandx{\improvement}[2][1=]{\todo[linecolor=Plum,backgroundcolor=Plum!25,bordercolor=Plum,#1]{#2}}
\newcommandx{\citationneeded}[2][1=]{\todo[linecolor=Aquamarine,backgroundcolor=Aquamarine!25,bordercolor=Aquamarine,#1]{#2}}
\newcommandx{\thiswillnotshow}[2][1=]{\todo[disable,#1]{#2}}

\usepackage[]{graphicx}
\graphicspath{{./figures/}{figures}}  
\usepackage{calc}  
\DeclareGraphicsExtensions{.pdf,.jpeg,.png,.jpg}

\usepackage{booktabs}
\usepackage[inline]{enumitem}
\newlist{mylist}{enumerate*}{1}
\setlist[mylist]{label=({\alph*})}

\usepackage{amsmath,amssymb}
\usepackage{longtable}
\usepackage{siunitx}
\usepackage{cleveref}
\Crefname{figure}{Fig.}{Figs.}
\usepackage{steinmetz}

\usepackage{array}

\usepackage[caption=false,font=normalsize,labelfont=sf,textfont=sf]{subfig}

\usepackage{fixltx2e}

\usepackage{stfloats}

\def\tsc#1{\csdef{#1}{\textsc{\lowercase{#1}}\xspace}}
\tsc{WGM}
\tsc{QE}

\newcommand{\ofk}{_k}
\newcommand{\params}{\boldsymbol\theta}
\newcommand{\hy}{\mathcal{H}}
\renewcommand{\vec}[1]{\mathbf{#1}}
\newcommand{\posegps}{\pose^\text{GNSS}}
\newcommand{\poseest}{\hat{\pose}}
\newcommand{\map}{\vec{m}}
\renewcommand{\lat}{\lambda}
\newcommand{\lon}{\phi}
\newcommand{\range}{\rho}
\newcommand{\bearing}{\beta}
\newcommand{\heading}{\psi}
\newcommand{\pose}{\vec{x}}
\newcommand{\observation}{\vec{z}}
\newcommand{\statobservation}{\vec{z}^{s}}
\newcommand{\obsstaticset}{\vec{Z}^{s}}
\newcommand{\landmrk}{{\vec{l}}}
\newcommand{\landmrkset}{\vec{L}}
\newcommand{\pihalf}{\frac{\pi}{2}}
\newcommand{\likelihoodf}{\mathcal{L}(\vec{z}\mid\pose_k,\map)}
\renewcommand{\comment}[1]{}
\newenvironment{psmallmatrix}{\left(\begin{smallmatrix}}{\end{smallmatrix}\right)}
\newcommand{\gaussian}[2]{\mathcal{N}\left({#1},{#2}\right)}
\newcommand{\var}[1]{\sigma_{#1}^2}

\newcommand{\res}{{\vec{r}}}
\DeclareMathOperator{\tr}{tr}

\begin{document}
	\let\WriteBookmarks\relax
	\def\floatpagepagefraction{1}
	\def\textpagefraction{.001}
	
	
	\shortauthors{Dagdilelis et al}
	
	\title [mode = title]{Cyber-resilience for marine navigation by information fusion and change detection}
	
	\author[AUT]{Dimitrios Dagdilelis}
	\cormark[1]
	\ead{dimda@elektro.dtu.dk}
	\credit{Methodology, Software, Writing}
	
	\affiliation[AUT]{organization={Automation and Control Group, Dept. of Electrical Engineering, Technical University of Denmark},
		addressline={Elektrovej B 326},
		city={Kgs. Lyngby},
		postcode={2800},
		country={Denmark}}
	
	\author[AUT]{Mogens Blanke}
	\ead{mb@elektro.dtu.dk}
	\credit{Conceptualization of this study, Methodology, Writing}
	
	\author[AUT]{Rasmus H. Andersen}
	\ead{rhjan@elektro.dtu.dk}
	\credit{Data collection, Software}
	
	\author[AUT]{Roberto Galeazzi}
	\ead{rg@elektro.dtu.dk}
	\credit{Conceptualization of this study, Methodology, Writing}
	
	\cortext[1]{Corresponding author}
	
	
	
	\begin{abstract}
		Cyber-resilience is an increasing concern in developing autonomous navigation solutions for marine vessels. This paper scrutinizes cyber-resilience properties of marine navigation through a prism with three edges: multiple sensor information fusion, diagnosis of not-normal behaviours, and change detection. It proposes a two-stage estimator for diagnosis and mitigation of sensor signals used for coastal navigation. Developing a Likelihood Field approach, a first stage extracts shoreline features from radar and matches them to the electronic navigation chart. A second stage associates buoy and beacon features from the radar with chart information. Using real data logged at sea tests combined with simulated spoofing, the paper verifies the ability to timely diagnose and isolate an attempt to compromise position measurements. A new approach is suggested for high level processing of received data to evaluate their consistency, that is agnostic to the underlying technology of the individual sensory input. A combined parametric Gaussian modelling and Kernel Density Estimation is suggested and compared with a generalized likelihood ratio change detector that uses sliding windows. The paper shows how deviations from nominal behaviour and isolation of the components is possible when under attack or when defects in sensors occur.
	\end{abstract}
	

	
	\maketitle
	
	\section{Introduction}\label{sec:intro}
	Safe navigation requires that a navigator is able to validate his position and heading at all times \citep{resolution1983}. With frequent attacks on satellite based navigation, autonomous vessels need automated means to validate essential navigation information and become resilient to attempts to misguide a vessel. 
	
	
	As the level of automation on board recreational and merchant vessels steadily increases, and solutions for autonomous marine navigation are emerging at the horizon, the number of attack surfaces are growing and concerns are raised about the security of ships against cyber-threats \citep{Felski2019}. The susceptibility of traditional ship navigation sensors to cyber-threats is well-known for the global positioning system (GPS) \citep{Humphreys2008a,Shepard2012,Kerns2014a,Bhatti2017a}, and automatic identification system (AIS), \citep{balduzzi2014a,iphar2015detection,goudossis2019towards}, but also ARPA Radar information, measured heading and other essential sensors can be compromised by an intruder \citep{Rugamer2015,Ioannides2016,Svilicic2020,balduzzi2014a,Katsilieris2013a}. The latter is possible if a malicious perpetrator is able to intelligently interfere with the navigation system either by exploiting IT-security vulnerabilities and deploy malicious software that alters internal information flow such as NMEA messages from sensors, or by leveraging security vulnerabilities in operational technology and counterfeit signals received by sensors -- e.g. global navigation satellite systems (GNNS) range and pseudo-measurements  -- and information systems -- e.g. AIS messages. 
	The purpose of such attacks could range from provoking collisions and groundings, to even hijacking of an autonomous vessel. Such subtle positioning control through live spoofing of the GNSS receiver has been demonstrated in the case of a surface vessel in \citep{Bhatti2017a} and in the case of an unmanned aircraft in \citep{Kerns2014a}. Systematic risks associated with autonomous marine vessels, their operation and control were discussed in \citep{Utne2019, Thieme2021}, and an architecture for risk mitigation was the subject of \citep{Dittmann2021a, Dittmann2021b}.
	
	Several approaches have been proposed in the direction of detecting and safeguarding against spoofing of the GNSS. They focus mainly on a sunder examination of the signal's electromagnetic characteristics \citep{wen2005a}, of the pseudo-range measurements \citep{Han2016b}, or on their coupling with an INS \citep{tan2018a,Liu2019,Tanil2016a,Xu2018,Broumandan2018,Li2013}. Spoofing and detection strategies were analyzed in~\citep{Psiaki2016}, including monitoring of the pseudo-ranges, fusion with an inertial measurement unit (IMU), and changes in the setup of the physical antenna. The detection of attacks on GNSS by means of sensor fusion with an IMU was addressed in~\citep{Kujur2020}. \cite{GrejnerBrzezinska2016} proposed the adoption of multi-sensor navigation systems to enhance the resiliency of the GNSS. \cite{Katsilieris2013a} demonstrated such approach for AIS monitoring by fusion of AIS information and radar measurements from land.
	
	In a marine navigation context, range and bearing of sets of objects was one of the classical techniques to obtain own position when coastline or navigation marks were in view, and this is a natural way to monitor integrity of a GNSS receiver. This paper proposes a generic approach that fuses measurements from the radar and information from the Electronic Navigation Chart (ENC) and derives the own ship's position estimate. 
	This approach abides with Safety of Life at Sea (SOLAS) regulations for knowledge of position, velocity and course of own ship at all times and it performs integrity monitoring of the vessel's position, heading and speed. Earlier results by the authors \citep{blanke2006, blanke2018} demonstrated that modelling using normal behaviours and analysis based on system structure provided a framework to isolate defect components, and \citep{Nissov2021} extended this to navigation in coastal waters using information from proprioceptive sensors (GNSS, compass, speed log) and exteroceptive sensors and information systems (ENC, ARPA radar, AIS). However, \cite{Nissov2021} did not show how the monitoring action could be achieved by unsupervised algorithms.
	
	This paper presents a condition monitoring system for the real-time supervision of own ship's position information based on an estimation-detection scheme. First, a two-stage estimator for coastal navigation is developed. Leveraging the Likelihood Field approach, the first stage extracts shoreline features from radar scans and matches them to the ENC to obtain the approximate position and heading of the vessel. Then the second stage refines such estimate by matching static landmarks (e.g. buoys) reported on the ENC with beacon features from the radar and performing trilateration on the matched landmarks. 
	Second, change detection algorithms based on the generalized likelihood ratio test (GLRT) are devised for both parametric and non parametric descriptions of the residual generated for the validation of the navigation sensor information. A combined monitoring approach using both a Gaussian GLRT and a Kernel Density Estimator (KDE) GLRT is then advocated to obtain detection that is robust for unsupervised monitoring in conditions where noise distributions are widely varying due to differing geometry of observed objects. The paper demonstrates the efficacy of the condition monitoring system on the detection of a GNSS spoofing attack employing full scale data collected during sea trials in the South Funen Archipelago (Denmark).

	\section{Problem formulation}\label{sec:Problem}
	A surface vessel (own ship) navigating in coastal waters is equipped with a standard commercial navigation sensor stack that include a GNSS receiver, a gyro compass and radars in S and X bands. Access to an ENC is also available. Own vessel can be subjected to a cyber-physical attack at any point in time.
	
	In the context of this article, it is assumed that it is the GNSS receiver being exposed to a cyber-physical attack, such as spoofing. Such an attack is subtle, in the sense that the attacker can inject slowly growing errors in the GNSS measurements and indirectly guide the vessel, through the autopilot, to an unwanted course and location. To be successful, such attack needs to happen without being noticed by the bridge (\cite{Bhatti2017a}). A cyber-physical attack needs to be detected and mitigated despite its incipient nature. This challenge is the subject of this research.
	
	As the vessel navigates through coastal waters, radar visible objects change. An additional difficulty of the problem is therefore to develop algorithms that are efficient under these conditions and be able to detect attempts to attack instruments with high probability of detection.

	\begin{table}[ptb]
		\renewcommand{\thetable}{\arabic{table}a} 
		\newcolumntype{L}[1]{>{\centering\arraybackslash}p{#1}}
		\renewcommand{\arraystretch}{1.3}
		\caption{List of variables - Part I}
		\label{tab:definitionsa}
		\centering
		\begin{tabular}{>{\centering\arraybackslash}m{0.13\columnwidth}m{0.7\columnwidth}}
			\toprule
			Variable                              & Definition                                                    \\ \midrule
			NED                                   & North-East-Down tangential plane                              \\
			$t$                                   & time                                                          \\
			$\{\}_k$                              & Subscript k denotes a variable at time $t_k$ \\
			$\pose$                               & Own ships pose                                                \\
			$\lat$                                & Latitude                                                      \\
			$\lon$                                & Longitude                                                     \\
			$\heading$                            & Heading    \\
			$\mu$ & mean value \\
			$\sigma$ & standard deviation \\
			$\observation$                        & Radar observation in range and bearing                      \\
			$\observation^{\text{geo}}$           & Radar observation in geodetic coordinates          \\
			$\range$                              & Range of a radar observation                                  \\
			$\bearing$                            & Bearing of a radar observation                                \\
			$\vec{p} $                            & Geodetic point                                                \\
			$\map$                                & Shoreline map                                                 \\
			$\mathcal{T}_{\vec{p}_0}(\cdot)$      & Geodetic to NED transformation                     \\
			$\mathcal{T}_{\vec{p}_0}^{-1}(\cdot)$ & NED to Geodetic transformation                     \\
			$\mathcal{L}(\cdot)$                  & Likelihood function                                           \\
			$\begin{psmallmatrix}
			p_{\text{hit}} \\
			p_{\text{random}} \\
			\range_{\text{max}} \\
			\sigma_{\text{LFM}}
			\end{psmallmatrix} $          & Likelihood Field Model parameters                             \\   \bottomrule
		\end{tabular}
	\end{table}
	
	\begin{table}[ptb]
		\addtocounter{table}{-1}
		\renewcommand{\thetable}{\arabic{table}b}
		\newcolumntype{L}[1]{>{\centering\arraybackslash}p{#1}}
		\renewcommand{\arraystretch}{1.3}
		\caption{List of variables - Part II}
		\label{tab:definitionsb}
		\centering
		\begin{tabular}{>{\centering\arraybackslash}m{0.1\columnwidth}m{0.8\columnwidth}}
			\toprule
			$\mathbf{Z}$                                                       & Set of radar observations $\observation$ forming a {360\textdegree} scan                            \\
			$p(\cdot)$                                                         & Probability Density Function (\emph{p.d.f.})                                                                  \\
			$\text{MLE}$                                                       & Maximum Likelihood Estimate                                                                                   \\
			$\statobservation$                                                 & A static landmark radar observation                                                                           \\
			${\statobservation}^{\text{NED}}$                                  & $\statobservation$ in NED coordinates                                                                         \\
			$\obsstaticset$                                                    & Set of detected static landmarks                                                                              \\
			$\landmrk$                                                         & A static landmark on the ENC                                                                                  \\
			${\landmrk}^{\text{NED}}$                                          & $\landmrk$ in NED coordinates                                                                                 \\
			$\landmrkset$                                                      & The set of available static landmarks on the ENC                                                              \\
			$\mathcal{L}$                                                      & Likelihood function                                                                                           \\
			$e(i)$                                                             & Association event between a landmark detected on the radar and a landmark on the ENC                          \\
			$\xi_{ij}$                                                         & Euclidean distance between $\statobservation_i , \landmrk_{j} $                                               \\
			$d_{ij}$                                                           & Euclidean distance between $\landmrk_i, \landmrk_j$                                                           \\
			$\Xi$                                                              & The set of pairwise euclidean distances                                                                       \\
			$r$                                                                & Range ratio                                                                                                   \\
			$a$                                                                & Angle of a vector from True North                                                                             \\
			$\hat{\vec{p}}_o$                                                  & Own ship's position estimate                                                                                  \\
			$ \hat{\vec{p}}_o^{\text{geo}} $                                   & Own ship's position estimate in geodetic coordinates                                                          \\
			$\hat{\psi}$                                                       & Own ship's heading estimate                                                                                   \\
			$\delta\psi$                                                       & Triangulated own ship's heading correction                                                                    \\
			$\vec{\Delta  p}_{\statobservation_{i}\rightarrow o}^{\text{NED}}$ & Triangulated own ship's position correction                                                                    \\
			\comment{$\mathcal{N}({\mu},\sigma)$                               & Gaussian distribution \emph{p.d.f.} with mean $\mu$ and variance $\sigma$                                     \\
				$ \mathcal{N}(\boldsymbol{\mu},\Sigma)$                            & Multivariate Gaussian distribution \emph{p.d.f.} with mean $\boldsymbol{\mu}$ and covariance matrix $\Sigma$} \\
			\bottomrule
		\end{tabular}
	\end{table}
	
	
	\subsection{Problem description}\label{ssec:Definitiions}
	When sailing at sea, it is demanded from the navigator to be at all times aware of the own ship's position, heading and velocity. Relying on the GNSS is common practice, but the specific source of information can be proven unreliable either because of  the sensor's malfunctioning, or because of an eminent cyber-attack (\cite{Humphreys2008a,Rugamer2015}). When GNSS usage is restricted, one can deploy exteroceptive sensors such as cameras, lidars, radars and sonars to localize the own ship. While for land vehicle navigation this is an extensively researched topic, in marine navigation and USV is has been only but relatively recently demonstrated in the work of \cite{Mullane2010,Han2015,han2016a,han2019,olofsson2020a}
	
	In our approach, through fusing processed information from different sensors, the Radar and the ENC, we present a position estimation process that does not rely on the GNSS. This independent position estimate allows us to subsequently perform monitoring of the integrity of the actual GNSS information and ultimately, provide an alarm in presence of a fault or a cyber-attack.
	
\begin{figure*}[ptb]
		\centering
		\includegraphics[width=0.8\textwidth]{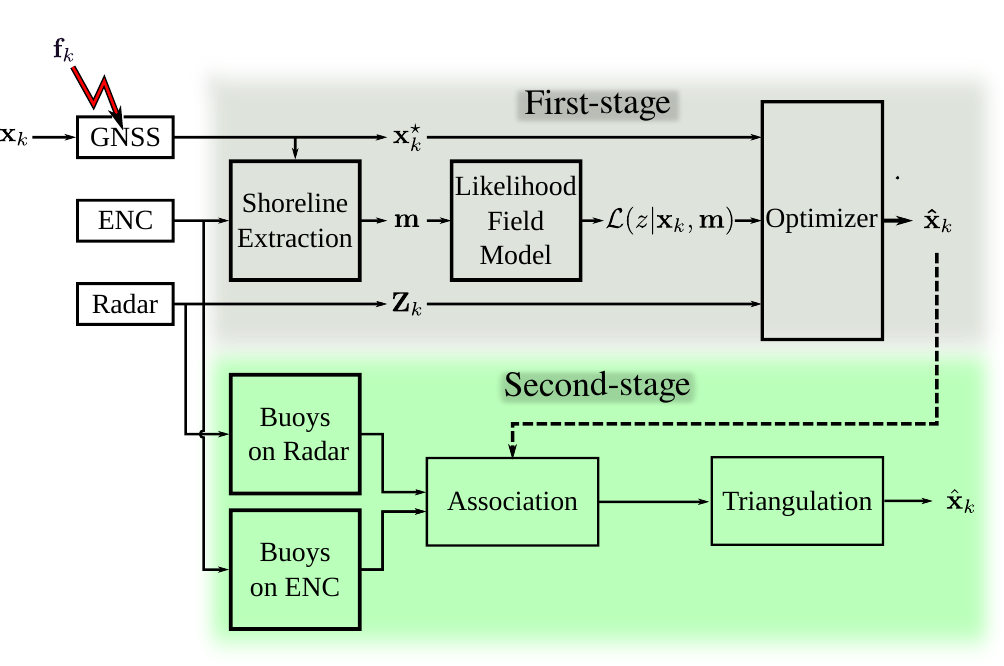}
		\caption{The figure illustrates the two-stage GNSSs independent positioning system. The first stage estimate is being used to solve the association problem in the second stage pose estimation.}
		\label{fig:two_stage2}
	\end{figure*}

	The proposed pipeline presented in the following subsections is a two-stage estimation process \Cref{fig:two_stage2} in which:
	\begin{itemize}
		\item The first stage provides a position estimate through probabilistically matching shoreline features from radar images to shoreline features on the ENC (see \Cref{fig:shoreline_extraction}).
		\item The second stage performs classical position estimation through landmark triangulation (\Cref{fig:triangulation}), as was a common navigation practice for centuries. We automate this process by using digitally detected features, namely buoys, beacons and navigation marks tracked on the radar. This step further refines the accuracy of the position estimate.
	\end{itemize}
	
	\begin{figure}[ptb]
		\centering
		\includegraphics[width=\columnwidth]{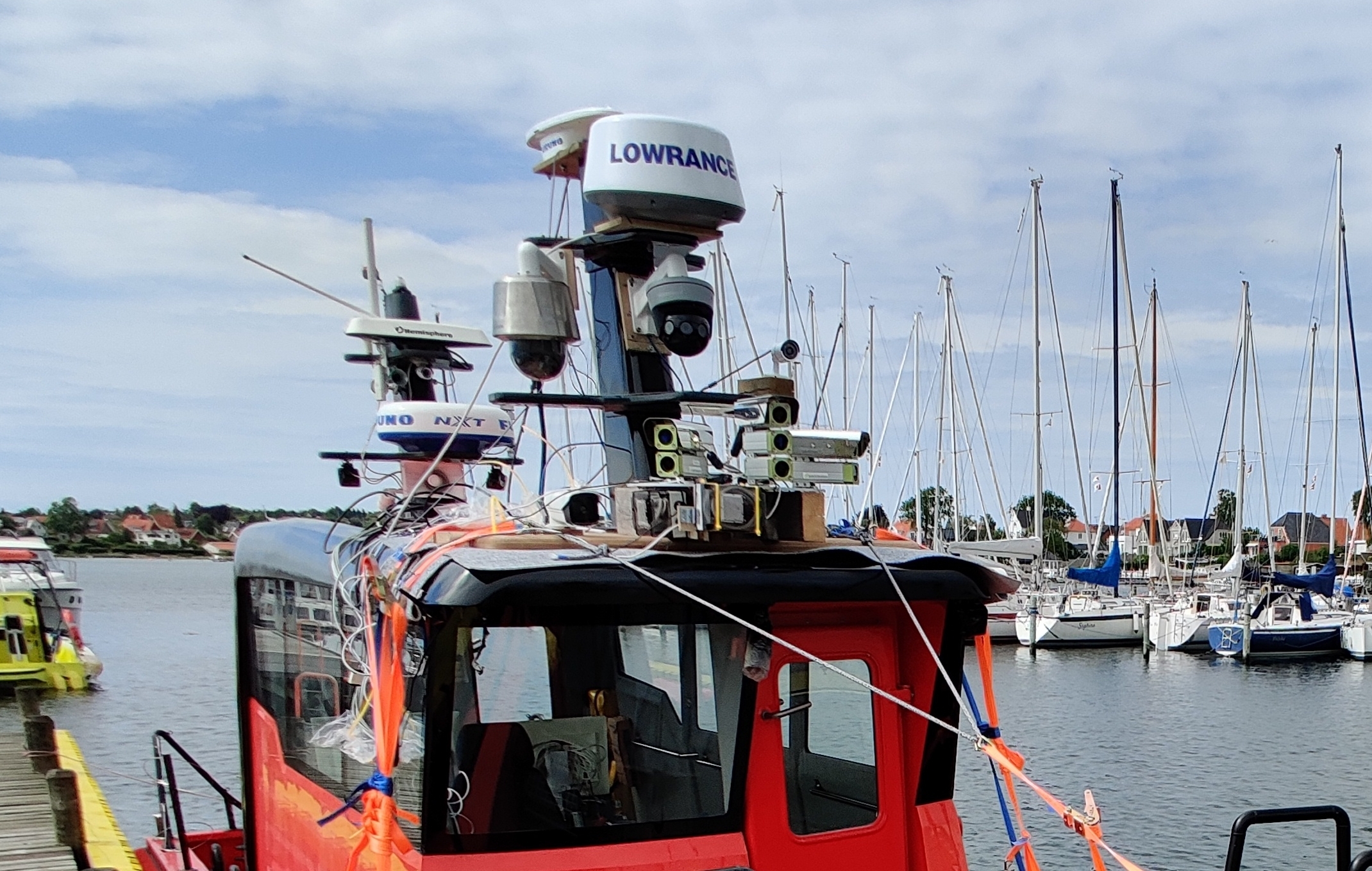}
		\caption{All data captured at sea. Mounted on the mast among other instrumentation one can spot the GNSS receiver and the X-band radar.}
		\label{fig:atsea}
	\end{figure}
	
	\section{Methods}\label{sec:Methods}
	\subsection{First-stage : Matching shoreline features between the radar and the ENC}
	\comment{
		
		A Terrain Aided Navigation (TAN) system is suitable for environments where reliability on standard navigation systems such as the GNSS can not be guaranteed (e.g., due to jamming or spoofing scenarios). Instead of relying on externally transmitted RF signals, TAN systems estimate the position of the own ship by correlating on-board exteroceptive sensor information with a given map of the terrain in the vicinity (\citep{Dunik2020}). Consequently, designing TAN systems with sensors that are inherently less prone to interference, such as vision or radar, unfolds a new potential in utilizing already existing methods as components of a cyber-resilient architecture.
		
		The structural analysis of a marine navigation system in the work of \cite{Nissov2021} has deliberated the cyber-resilience potential of a TAN system based on a radar sensor and the Electronic Navigational Charts(ENC). Therefore in this work we present our approach in the design of such a system.
	}
	
	Similarly to a Terrain Aided Navigation system, we hereby propose an estimation process that is probabilistically matching extracted shoreline-features from the radar scans, to expected shoreline features on the ENC. The strength of our first-stage approach is that
	\begin{itemize}
		\item It does not depend on any form of any data association between the sensor and the map.
		\item While not explicitly solving the global localization problem, it can recover the own ships pose, with only but a rough idea of the sailing location.
		\item Despite the long range radar is a noisy sensor, the proposed estimator remains robust to false detection on the radar.
		\item Is crucial in solving the data-association problem in the second-stage.
	\end{itemize}
	\begin{figure}[ptb]
		\centering
		\includegraphics[width=\columnwidth]{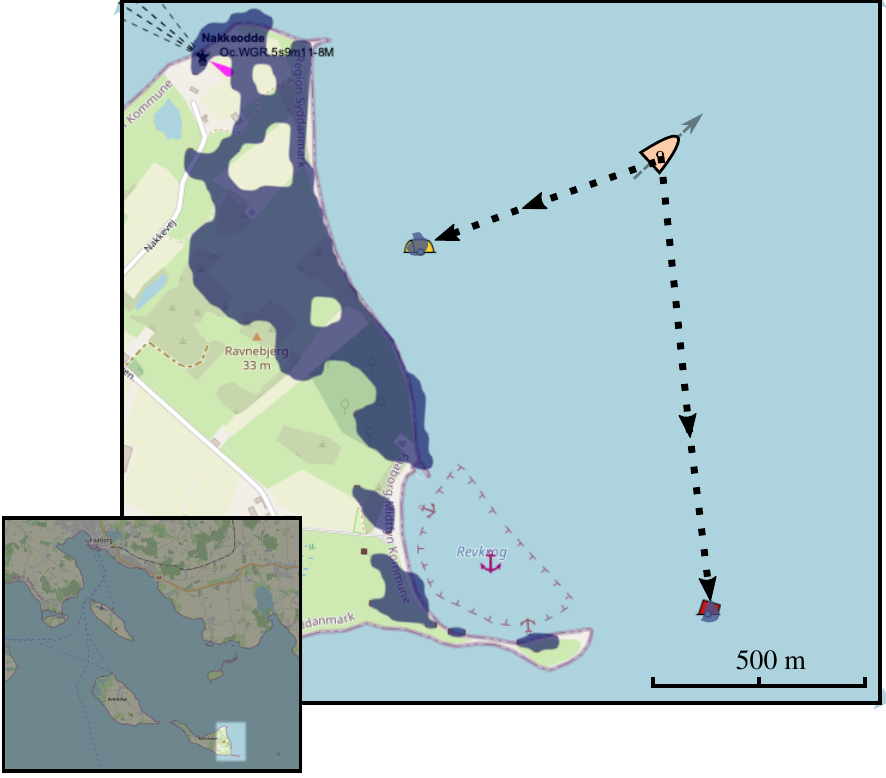}\hspace*{0.5cm}
		\caption{Two-step localization. First, a radar scan is matched to the shoreline, then the detected buoys are associated to the buoys on the ENC, thus delivering the final pose estimate.}
		\label{fig:two_stage1}
	\end{figure}

	\subsubsection{Shoreline Extraction}
	
	\paragraph{Own ship pose}
	Let $\pose_k = \begin{psmallmatrix} \lat & \lon & \psi \end{psmallmatrix}^T $ represent the 3-DOF pose of the own ship relative to a given ENC map $\map$. Where $\lat, \lon, \psi$ correspond to standard notation for latitude, longitude and heading from true north respectively at the time index $t_k$. Additionally, let $\observation = \begin{psmallmatrix} \range & \bearing \end{psmallmatrix}^T$ represent a shoreline radar observation in range and bearing, where bearing is the clockwise angle from the direction of heading. Then a full 360 degree radar scan is defined as the set $\mathbf{Z}_k$. It is assumed that the heading of the own ship during the radar's rotation is either constant, or known 
	
	\paragraph{Shoreline-feature extraction}
	Let the zero-altitude geodetic coordinate domain
	
	\begin{eqnarray}
	\mathcal{A} \triangleq \{\vec{p}= (\lat, \lon) \in \mathbb{R}^2 \mid \lambda \in \begin{bmatrix}
	-\pihalf,\pihalf
	\end{bmatrix} \land \phi \in \begin{bmatrix}
	-\pi,\pi
	\end{bmatrix}\}
	\end{eqnarray}
	Then let the shoreline extraction mapping $\mathcal{F}:\mathcal{A} \to \mathcal{A}$ represent the set of geodetic points that correspond to shorelines on the ENC.

	\paragraph{Minimum Euclidean distance transformation}
	Let $ \mathcal{D}(\vec{p},\vec{m}) $ represent the minimum Euclidean distance of a geodetic point $\vec{p} \in \mathcal{A}$ to any point on the shoreline map $\map = \mathcal{F}(\mathcal{A})$
	
	\begin{equation}
	\mathcal{D}(\vec{p},\map) =  \min{\| \mathcal{T}_{\vec{p}_0}(\vec{p})\|}, \, \forall \vec{p}_0 \in \map
	\end{equation}
	
	where $\mathcal{T}_{\vec{p}_0}(\cdot)$ is the coordinate transformation from the geodetic to a North-East sea-level (zero altitude) tangential reference system centered at $\vec{p}_0=\begin{psmallmatrix}
	\lat_0 & \lon_0
	\end{psmallmatrix}^T$  and $\mathcal{T}_{\vec{p}_0}^{-1}(\cdot)$ the corresponding inverse transformation (\cite{Nerem2001}).
	\begin{eqnarray}
	\vec{p}^{\text{NED}}= \mathcal{T}_{\vec{p}_0}(\vec{p}^{\text{geo}})={\begin{bmatrix}0&M(\phi_0)\\N(\phi_0)\cos \phi_0 & 0\end{bmatrix}}(\vec{p}^{\text{geo}}-\vec{p}_0) \\
	\vec{p}^{\text{geo}}=\mathcal{T}^{-1}_{\vec{p}_0}(\vec{p}^{\text{NED}})={\begin{bmatrix}0 & \frac{1}{M(\phi)}\\ \frac{1}{N(\phi_0)\cos \phi_0} & 0 \end{bmatrix}}\vec{p}^{\text{NED}}+\vec{p}_0\\
	\end{eqnarray}
	Where the WGS84 model and its parameters are
	\begin{eqnarray}
	N(\phi )={\frac {a^{2}}{\sqrt {a^{2}\cos ^{2}\phi +b^{2}\sin ^{2}\phi }}}={\frac {a}{\sqrt {1-e^{2}\sin ^{2}\phi }}}\\
	M(\phi )={\frac {a\left(1-e^{2}\right)}{\left(1-e^{2}\sin ^{2}\phi \right)^{\frac {3}{2}}}}\\
	e^{2}=1-{\frac {b^{2}}{a^{2}}}
	\end{eqnarray}
	\subsubsection{Likelihood Field Model}
	The likelihood $\mathcal{L}$ of a shoreline radar observation $\observation$, given the own ship's pose $\pose_k$ and the shoreline map $\map$  is defined as \addtocounter{equation}{-1}
	\begin{subequations}
		\label{eq:likelihoodfunction}
		\begin{eqnarray}
		\mathcal{L}(\observation\mid \pose_k, \map) &= p_{hit} \cdot \frac{1}{2\pi \sigma_{\text{LFM}}}e^{-\frac{1}{2}(\frac{\mathcal{D}(\observation^{\text{geo}}, \map)}{\sigma_{\text{LFM}}})^2}  \label{eq:sub1}\\
		&+ p_{random}\frac{1}{\range_{max}} \label{eq:sub2}
		\end{eqnarray}
	\end{subequations}
	where $ \observation^{\text{geo}}\in \mathcal{A} $ is a radar observation transformed from NED to geodetic coordinates
	\begin{equation}\label{eq:radar_geo}
	\observation^{\text{geo}} = \mathcal{T}_{\pose_k}^{-1}(  \range \phase{\bearing + \heading_k})
	\end{equation}
	The likelihood function in \eqref{eq:likelihoodfunction} is comprised of two terms. The term in \eqref{eq:sub1} models the  likelihood of a radar observation $\observation^{\text{geo}}$ given its distance to the closest point on the shoreline map $\map$, while \eqref{eq:sub2} models the likelihood of a clutter measurement by means of an  uniform distribution across the radar's maximum spoke range $\range_{\text{max}}$.
	
	The model parameters $p_{\text{hit}}, \sigma_{\text{LFM}}, p_{\text{random}}$ are being estimated through Expectation Maximization on a ground truth data-set (\cite{Thrun2001})where $\pose_k , \mathbf{Z}_k, \map$ are known. In absence of actual ground-truth information while out at sea, we used the GNSS position as ground-truth.
	
	\paragraph{Maximum-likelihood estimate}
	
	What we have established in the previous paragraph, is the likelihood function of an underlying shoreline classification model, in the form of global nearest-neighbor soft binary classifier, with the predictors being the own ship's pose $\pose_k$ and a radar scan  $\mathbf{Z}_k$.
	
	An estimate of the own ship's pose $\vec{x}_k$ can be obtained as the parameter that maximizes the conditional \emph{p.d.f}  $p(\pose_k\mid\vec{Z}_k,\vec{m})$
	\begin{eqnarray}
	\hat{\pose}_k=  \text{MLE} \ \vec{x}_k =\displaystyle \underset {\mathbf{x}_k }{\operatorname {arg\;max}\ }p(\pose_k\mid\vec{Z}_k,\vec{m})  
	\end{eqnarray}
	
	Assuming that radar observations $\observation \in \vec{Z}_k$ are independent of each other, then
	\begin{eqnarray}
	\mathcal{L}(\vec{Z}_{k}\mid\pose_k,\vec{m}) = \prod_{\observation \in \vec{Z}_{k}}\mathcal{L}(\observation \mid \pose_k, \vec{m} )
	\end{eqnarray}
	By Bayes rule:
	\begin{eqnarray}\label{eq:bayes}
	p(\pose_k\mid\vec{Z}_k,\vec{m}) = \frac{\mathcal{L}(\vec{Z}_{k}\mid\pose_k,\vec{m}) p(\pose_k) }{\int_{\mathcal{A}}\mathcal{L}(\vec{Z}_{k}\mid\pose_k,\vec{m}) p(\pose_k)d\pose_k}
	\end{eqnarray}
	The own ship's pose $\vec{x}_k$ is being marginalized out in the denominator of  \eqref{eq:bayes}. Additionally, by using an uninformative uniform prior pose distribution
	\begin{eqnarray}\label{eq:bayesprior} p(\vec{x}_k)=\text{const}
	\end{eqnarray}
	From \eqref{eq:bayes}\eqref{eq:bayesprior} and because of the logarithm being a strictly monotonic function, one can obtain a maximum likelihood estimate of $\vec{x}_k$ by means of maximizing the overall log-likelihood function
	\begin{eqnarray}
	\log p(\pose_k\mid\vec{Z}_k,\vec{m}) \sim \log \mathcal{L}(\vec{Z}_{k}\mid\pose_k,\vec{m}) p(\pose_k)
	\end{eqnarray}
	\begin{eqnarray}
	\hat{\pose}_k=\displaystyle \underset {\mathbf{x}_k }{\operatorname {arg\;max}\ }p(\pose_k\mid\vec{Z}_k,\vec{m}) =\\
	=\displaystyle \underset {\mathbf{x}_k }{\operatorname {arg\;max} }\,\log \mathcal{L}(\vec{Z}_{k}\mid\pose_k,\vec{m}) \label{eq:est}
	\end{eqnarray}
	Given a GNSS measurement $\pose^{\text{GNSS}}_k=\begin{psmallmatrix} \lat & \lon & \heading \end{psmallmatrix}^T$, we introduce the error estimate $\hat{\delta \pose}$ such that
	\begin{eqnarray}
	\hat{\pose}_k = \pose^{\text{GNSS}}_k + \hat{\delta \pose}\\
	\hat{\delta \pose } = \displaystyle \underset {\delta \pose }{\operatorname {arg\;max} }\,\log \mathcal{L}(\vec{Z}_{k}\mid \pose^{\text{GNSS}}_k +\delta\pose,\vec{m})\label{eq:mledx}
	\end{eqnarray}
	
	The error's likelihood function $\log \mathcal{L}(\vec{Z}_{k}\mid \pose^{\text{GNSS}}_k +\delta\pose,\vec{m})$ in \eqref{eq:mledx} is plotted in \Cref{fig:likelihood_function}. In nominal error-free conditions, the observations from the radar should match the reported pose from the GNSS and the map, resulting in an error likelihood function peaked around $\delta \pose \approx \vec{0}$.

\begin{figure}[ptb]
		\centering
		\includegraphics[width=2.5in]{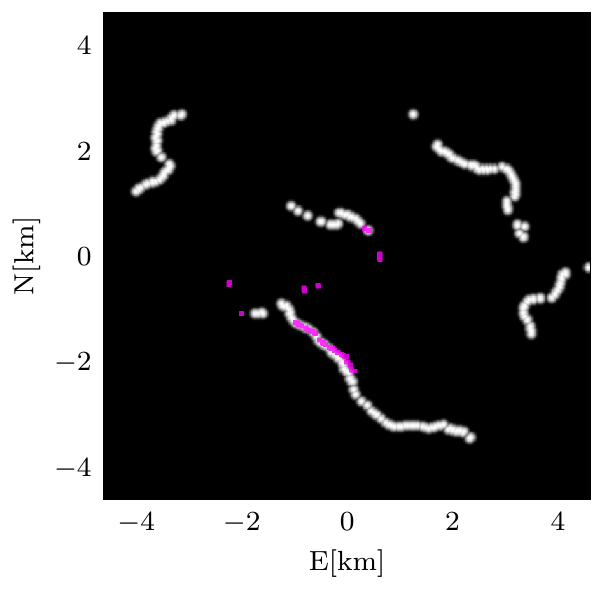}\hspace*{0.5cm}
		\caption{The figure displays in grayscale the likelihood function $\likelihoodf$ over $z$. In magenta the radar-observed shoreline from scan $\mathbf{Z}_k$.}
		\label{fig:shoreline_extraction2}
	\end{figure}

\begin{figure}[ptb]
		\centering
		\subfloat[][Marginal log-likelihood function,
		$
		\delta\pose= \begin{psmallmatrix}
		\delta N , \delta E , 0
		\end{psmallmatrix}^T$
		\label{subfig-1:dummy}]{%
			\includegraphics[width=3.2in]{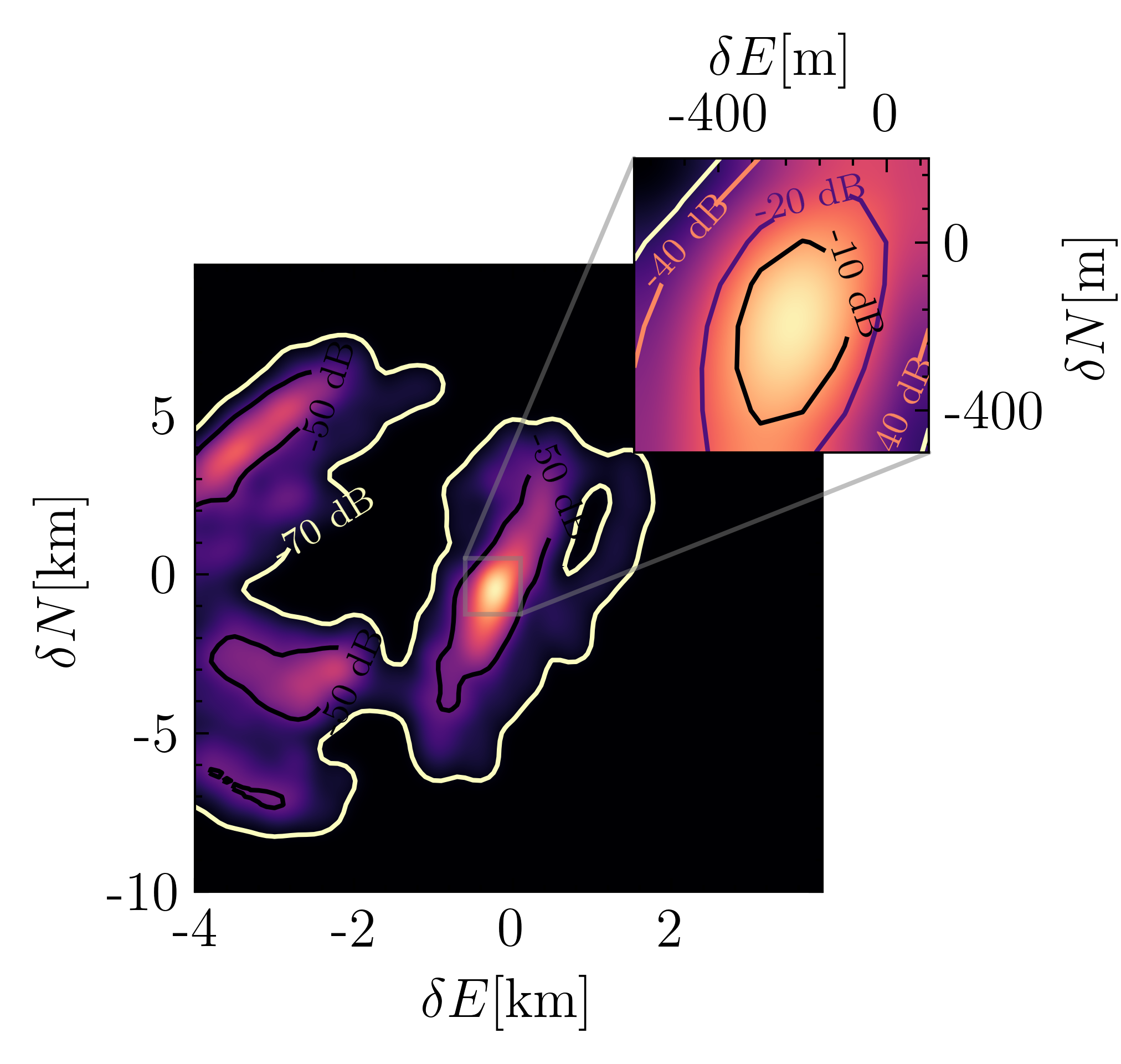}
		}
		\hfill
		\centering
		\subfloat[][Marginal log-likelihood function,
		$\delta\pose= \begin{psmallmatrix}
		\delta N , 0 , \delta{\psi}
		\end{psmallmatrix}^T$
		\label{subfig-2:dummy}]{%
			\includegraphics[width=1.8in]{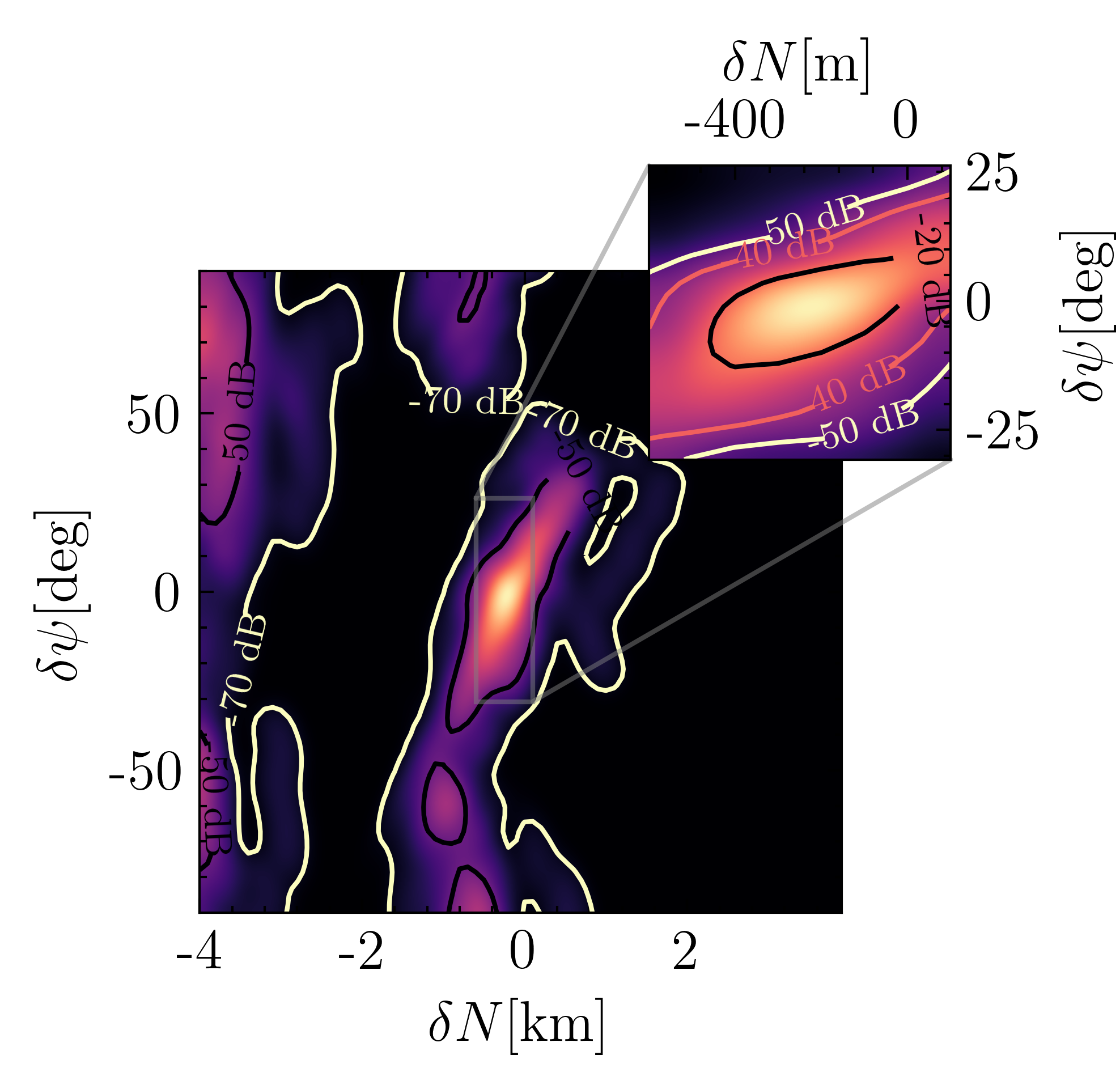}
		}
		\hfill
		\centering
		\subfloat[][Marginal log-likelihood function, $\delta\pose= \begin{psmallmatrix}
		0 , \delta E , \delta\heading
		\end{psmallmatrix}^T$\label{subfig-3:dummy}]{%
			\includegraphics[width=1.8in]{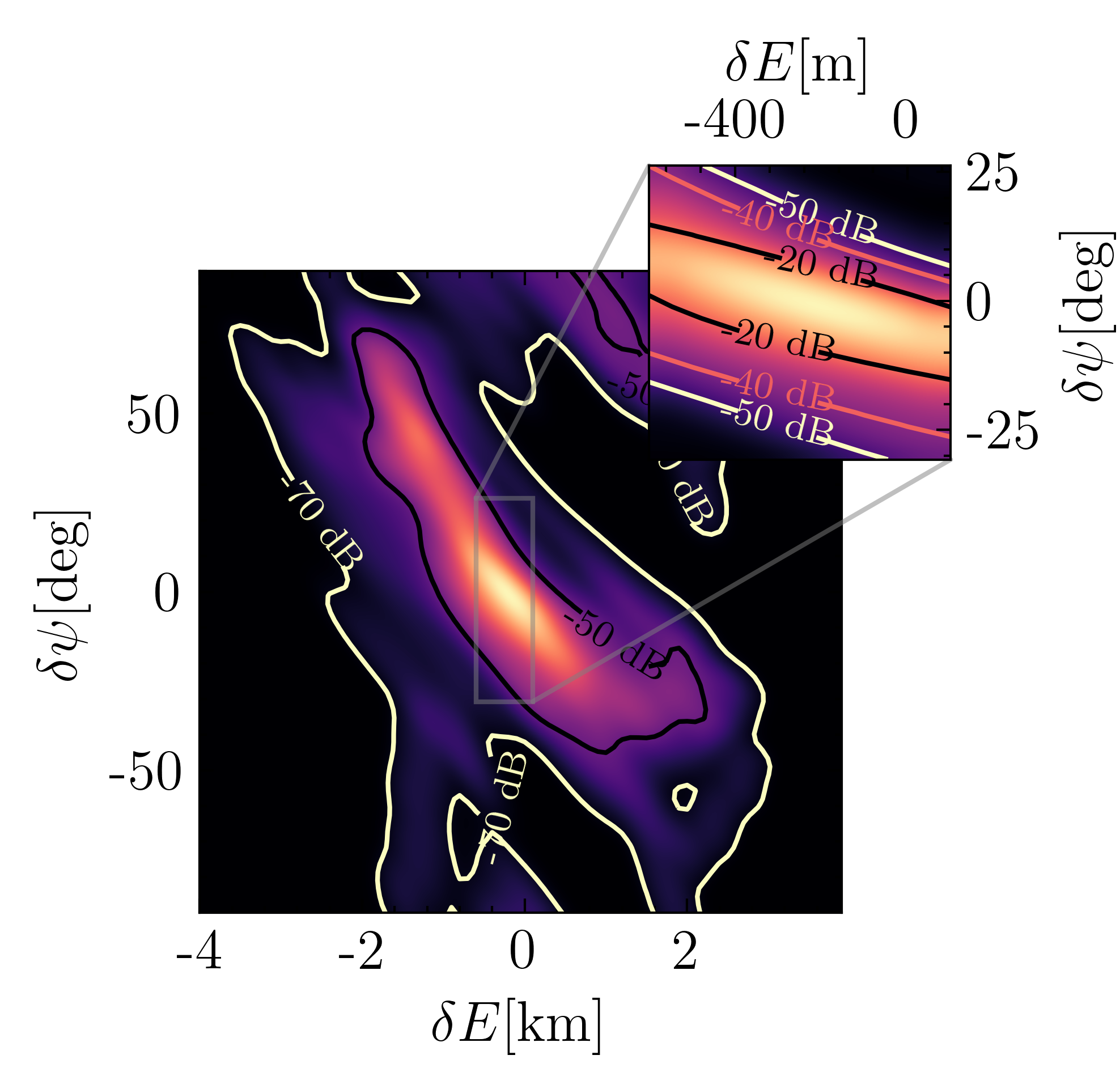}
		}
		
		\caption[]{The figures display the log-likelihood function of a radar scan $\vec{Z}_{k}$  given a map $\map$ $\log \mathcal{L}(\vec{Z}_{k}\mid \pose_k ,\vec{m})$ evaluated around $\pose^{\text{GNSS}}$  (see also \Cref{fig:shoreline_extraction2}).
			In absence of an error in the GNSS measurement, we observe the the global maximum of the error function to be found around $\delta\pose=\vec{0}$. 
		}
		\label{fig:likelihood_function}
	\end{figure}

	The major advantages of using the likelihood-field model over other existing (\citep{burgard2006a,pfaff2007a}) beam-based models , are
	\begin{enumerate}[label=(\alph*)]
		\item The smoothness of the likelihood function  (see \Cref{fig:likelihood_function}) $\mathcal{L}(\vec{Z}_k | \pose_k, \vec{m})$ (\citep{Thrun2001}) which has been observed to lead to consistent convergence of the numerical optimizer even for initial search points distant from the global maximum. This can be verified in \Cref{fig:likelihood_function}, which is an example of the error log-likelihood function corresponding to the extracted shoreline features of the example illustrated in \Cref{fig:shoreline_extraction,fig:shoreline_extraction2}.
		
		\item The robustness to outliers, the shoreline features are sensitive to land morphology, sensor noise and even temporary environmental factors such as tides. The likelihood field naturally rejects false outlying shoreline feature points from the radar.
		\item Does not require solving the data association between the radar and the ENC information, and thus can be used as the first building block in a pipeline used to recover the own ship's pose, that requires only crude geographical knowledge of where the own ship is located.
		\item The pose recovered from this process, will be used as prior knowledge in the static landmark localization algorithm described in the following section, where a relatively accurate initial pose is required to solve the data association problem.
		
	\end{enumerate}
	
	\begin{figure}[ptb]
		\centering
		\includegraphics[trim=0cm 0 0 0cm]{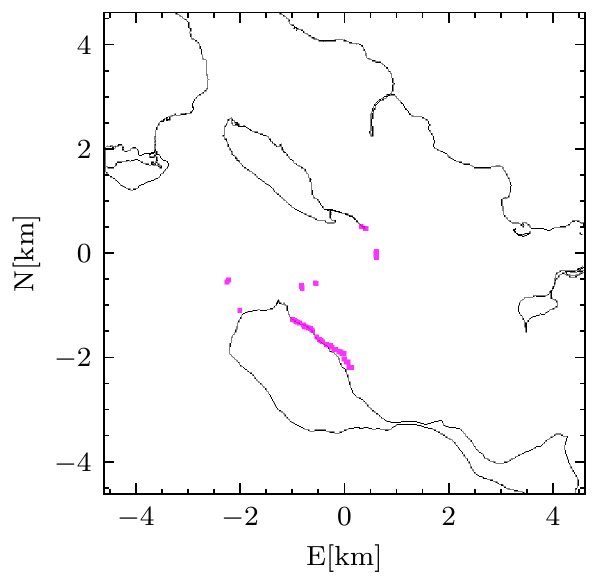}\hspace*{0.5cm}
		\caption{In black the ENC extracted shoreline, in magenta the radar closest range measurements across every spoke.}
		\label{fig:shoreline_extraction}
	\end{figure}

	\comment{
		\begin{figure}
			\centering
			\includegraphics[width=2.5in]{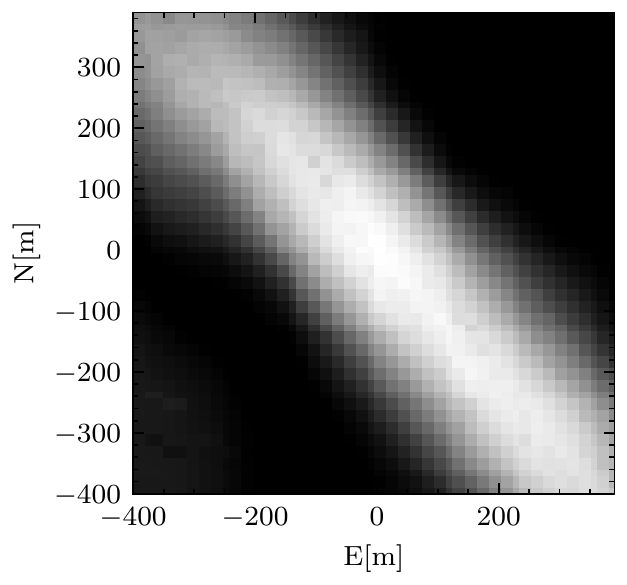}
			\caption[]{The figure displays in gray-scale the overall log-likelihood function $\prod_{\observation \in \vec{Z}_{k}} \likelihoodf$ over the own ship's pose
				$\pose_k = \mathcal{T}_{\vec{p}_0}^{-1}\big(
				\begin{psmallmatrix}N\\E\end{psmallmatrix},
				\begin{psmallmatrix}\lat\\\lon\end{psmallmatrix}
				\big)
				$
				$
				\pose_k = \begin{psmallmatrix}
				\lat \\ \lon \\ \heading
				\end{psmallmatrix}_{\text{GNSS}}
				+
				\begin{psmallmatrix}
				\text{N} \\ \text{E} \\ 0
				\end{psmallmatrix}
				$.}
			\label{fig:shoreline_extraction}
		\end{figure}
	}
	
	\subsubsection{Optimizer}
	
	In general we obtain a multi-modal likelihood function for the numerical optimization problem of \eqref{eq:est}, hence we avoid commonly applied descent methods. Instead we deploy Particle Swarm Optimization (PSO), a heuristic evolutionary optimization algorithm that is simple, effective and computationally efficient algorithm. Additionally, it has, through our validation on real-data, proven to provide robust and reliable solutions. An analysis to the solution of the optimization problem is beyond the scope of our work. The reader is referred to \cite{Marini2015,Zhang2019,Kulkarni2011,John2019} for a description and applications of PSO in localization problems.
	
	\subsection{Second-stage : Static-landmark-matching}
	
	It is an every-day navigators skill and task, being able to estimate the own ship's position based on observed landmarks on the horizon, usually measuring their relative bearings using an alidade and associating them with known-features on the map. We present an automated \emph{detect and associate} process, that first detects buoy features on radar scans and measures their relative range and bearing, then it associates them with the ENC and lastly solves a combined triangulation-trilateration system of equations to accurately estimate the own ship's position and heading.
	
	\subsubsection{Buoys on the radar scans}
\begin{figure}
		\centering
		\includegraphics[width=2.5in]{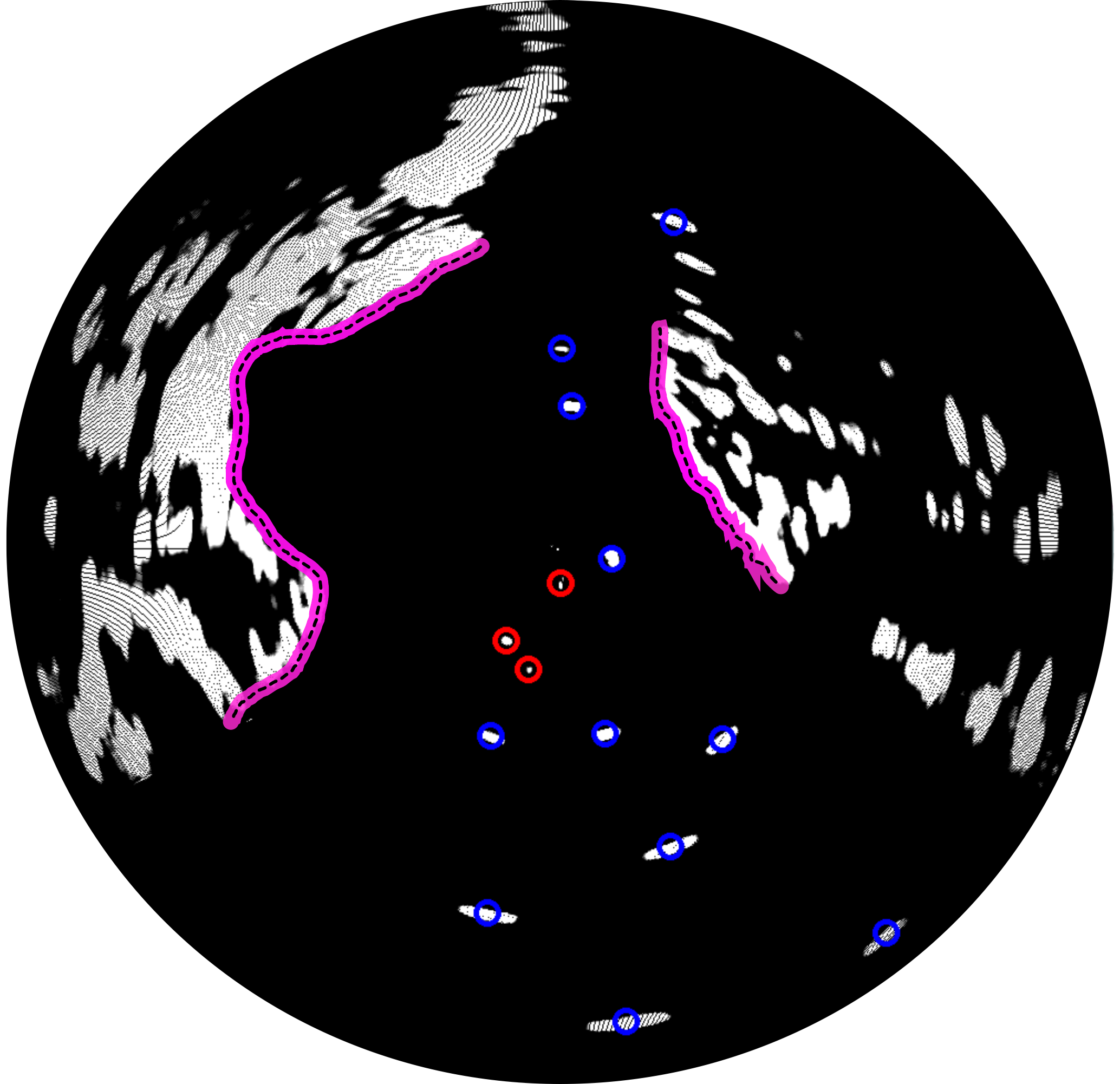}
		\caption[]{The figure displays different level feature extraction from a radar scan $ \vec{Z}_k $. In magenta the extracted shoreline. Circles correspond to static tracked targets. Target's with zero velocity are assumed to be buoys  $\obsstaticset$ . Blue targets are confirmed because the exist in the previous scans  $\vec{Z}_{k-n}, n=1,2,3$, red are unconfirmed.}
		\label{fig:buoys_feature_tracking}
	\end{figure}

	A full radar scan can provide more information than just a contour of the surrounding shoreline (\Cref{fig:buoys_feature_tracking}). By leveraging machine-learning methodology, we extract features from the radar-data in order to perform landmark-based localization. The features are the centroids of clusters within a scan, which are subsequently being tracked through successive radar scans, using a Probabilistic Data Association Filter and an Extended Kalman Filter (\cite{Bar-Shalom2009}). The EKF assumes the target dynamics following a constant turn rate and velocity magnitude model (\cite{Yuan2014a}). This framework provides an estimate of the velocities of the tracked features. Features with zero velocity are assumed to be static landmarks, i.e.\ being anchored ships, buoys or obstacles. In practice it is very rare that we consistently detect and track a static target which classifies to an object different than a buoy. Additionally, buoys at sea usually carry on a radar reflector, which is making them highly visible on the radar.

	Following the previous section's convention, let  $\obsstaticset_k$ be the set of $n$ static landmarks as tracked from the radar at time instant $t_k$. The index $k$ is dropped for clarity.
	\begin{eqnarray}
	\obsstaticset = \{\displaystyle \statobservation_{1},\statobservation_{2},\ldots ,\statobservation_{n} : \vec{\statobservation}_{i} = \left(\rho_i, \beta_i\right) \}
	\end{eqnarray}
	\subsubsection{Buoys on the ENC}
	Let the set of all landmarks $\landmrkset$ within the range of detection of the own ship
	\begin{eqnarray}
	\landmrkset=\{\displaystyle \landmrk_{1},\landmrk_{2},\ldots ,\landmrk_{n} : \landmrk_{i} \in \mathcal{A} \}
	\end{eqnarray}
	Such information is readily available from digital files.
	\subsubsection{Association}
\begin{figure*}
		\centering
		\includegraphics[width=0.8\textwidth]{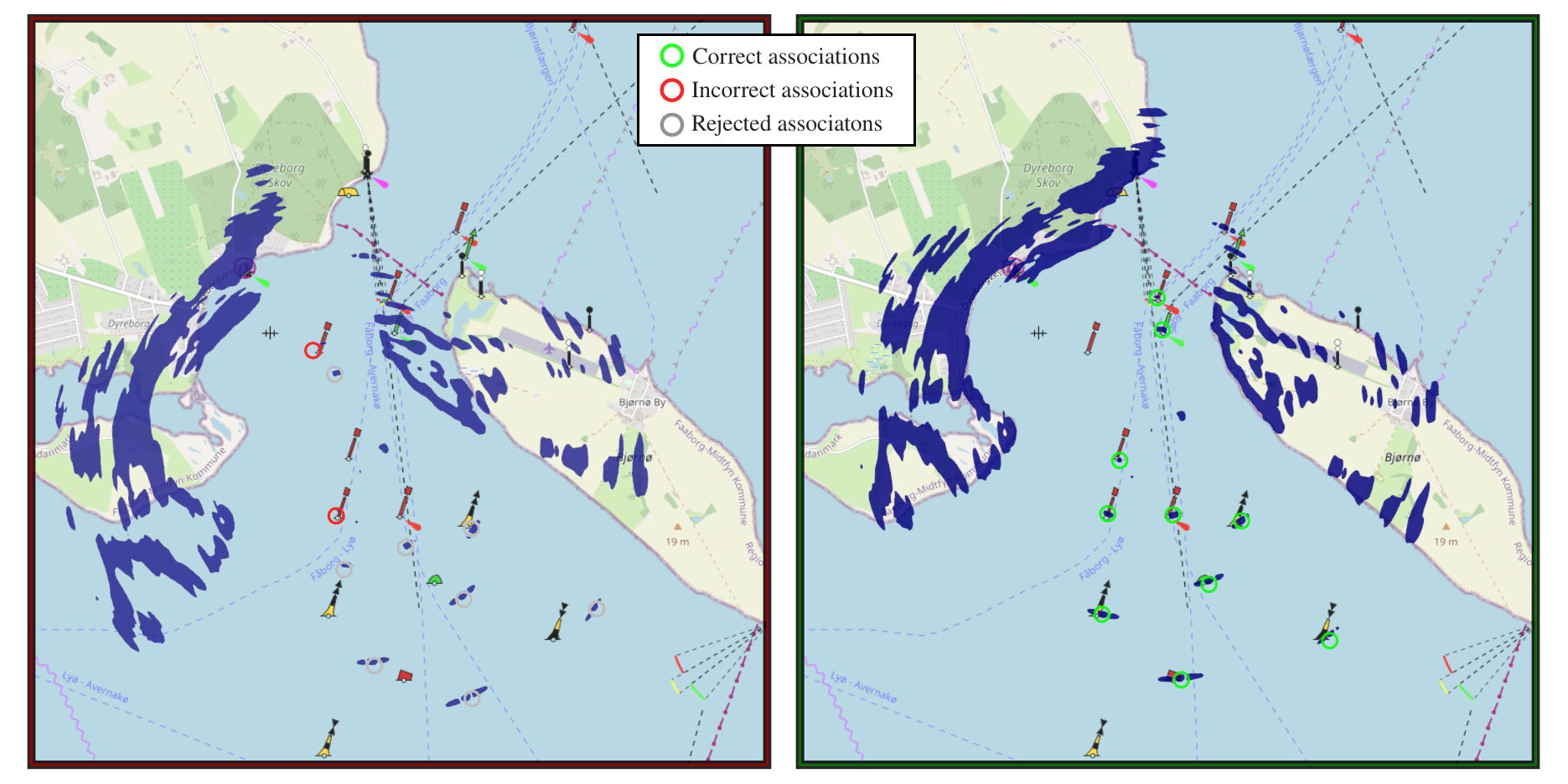}
		\caption{Left hand side, incorrect associations because of an error in the pose estimate $\hat{\pose}_k$. Right hand side, the position estimate is updated from the first-stage estimator, now the buoys on the map can be easily associated to buoys on the radar. }
		\label{fig:association}
	\end{figure*}

	In a cluttered marine environment, the received measurements may not all arise from the real targets. Some of them may be from clutter or false detections, for instance an anchored ship. As a result, there always exist ambiguities in assigning detected buoys on the radar to buoys on the ENC. In the left figure of \Cref{fig:association}, we illustrate an example of false associations, where we can observe from the general misalignment of the radar image to the ENC, that there is an error in the own ship's position and heading. This error is leading to false buoy associations between what we observe and what exists in the world. In the right figure of \Cref{fig:association}, after roughly aligning the radar scan to the map, the problem greatly simplifies and leads to correct associations. This is where the first-stage estimator proves extremely valuable. In presence of an error in the GNSS measurements, the first-stage shoreline feature matching estimator, is  providing this initial alignment.

	The association problem is that of identifying a one-to-one correspondence between the sets $\obsstaticset, \landmrkset$. The association can be formally represented by the discrete mapping $e(i)$, which associates a static target $\statobservation_i$ to the landmark $\landmrk_{e(i)}$
	\begin{eqnarray}
	e(i) = \displaystyle \underset {j}{\operatorname {arg\;min} }\, (\xi_{ij}) \mid \xi_{ij} \in \Xi_i\\
	\Xi_i = \{\xi_{ij} :  \xi_{ij} > \delta_{\text{min}} \} \\
	\xi_{ij} = |{\statobservation}^{\text{NED}}_{i} - {\landmrk}^{\text{NED}}_{j}|
	\end{eqnarray}
	The approach presented above, is in practice a Global Nearest Neighbor classification, or else a k-nearest neighbor association with $k=1$.
	
	Given the set $L^{\star}$ of two associated pairs
	\begin{eqnarray}
	L^{\star}=\{\alpha_{i,j}\}=\{\{\statobservation_i, \landmrk_{e(i)}\},\{\statobservation_j, \landmrk_{e(j)}\}\}=\\
	=\{\{\statobservation, \landmrk\}_{i,j}\}
	\end{eqnarray}
	where for naming simplicity, it is assumed that $e(i)=i, e(j)=j$

	Then the range ratio $r_{ji}$ and bearing difference $\beta_{ji}$ are defined respectively
	\begin{eqnarray}
	r_{ji} &= \frac{\rho_j}{\rho_i} \label{eq:rangeratio} \\
	\beta_{ji} &= \beta_j - \beta_i \label{eq:bearingdiff}
	\end{eqnarray}

	\begin{figure}[ptb]
		\centering
		\includegraphics[width=2in]{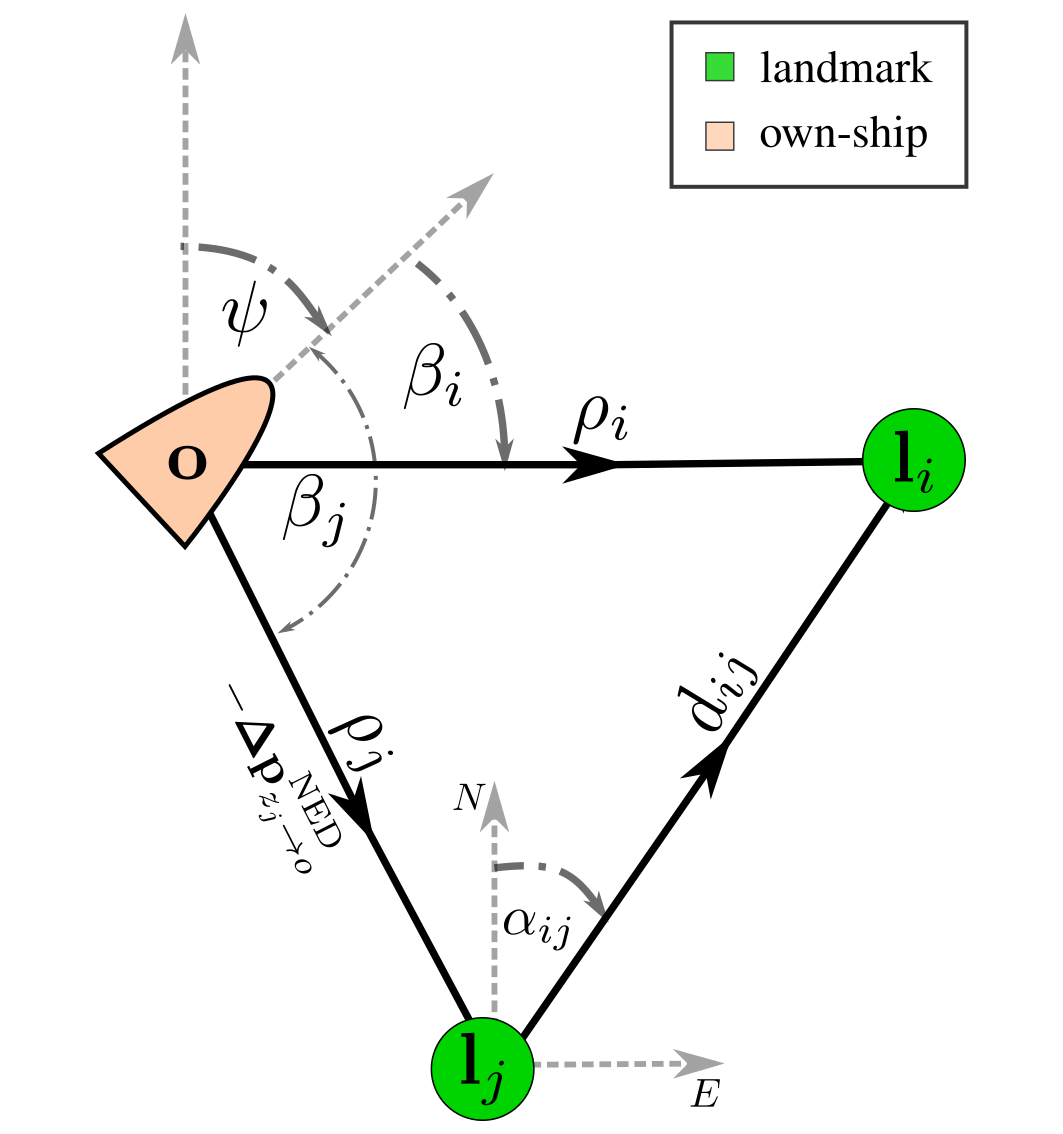}
		\caption{Own ship pose estimation, by resection of two observed landmarks $\landmrk_{i,j}$, whose position is known on the ENC. The own ship is able to measure the landmark's relative range and bearing information $\observation_{i,j}$}.
		\label{fig:triangulation}
	\end{figure}

	\begin{figure}[ptb]
		\centering
		\includegraphics[width=2.5in]{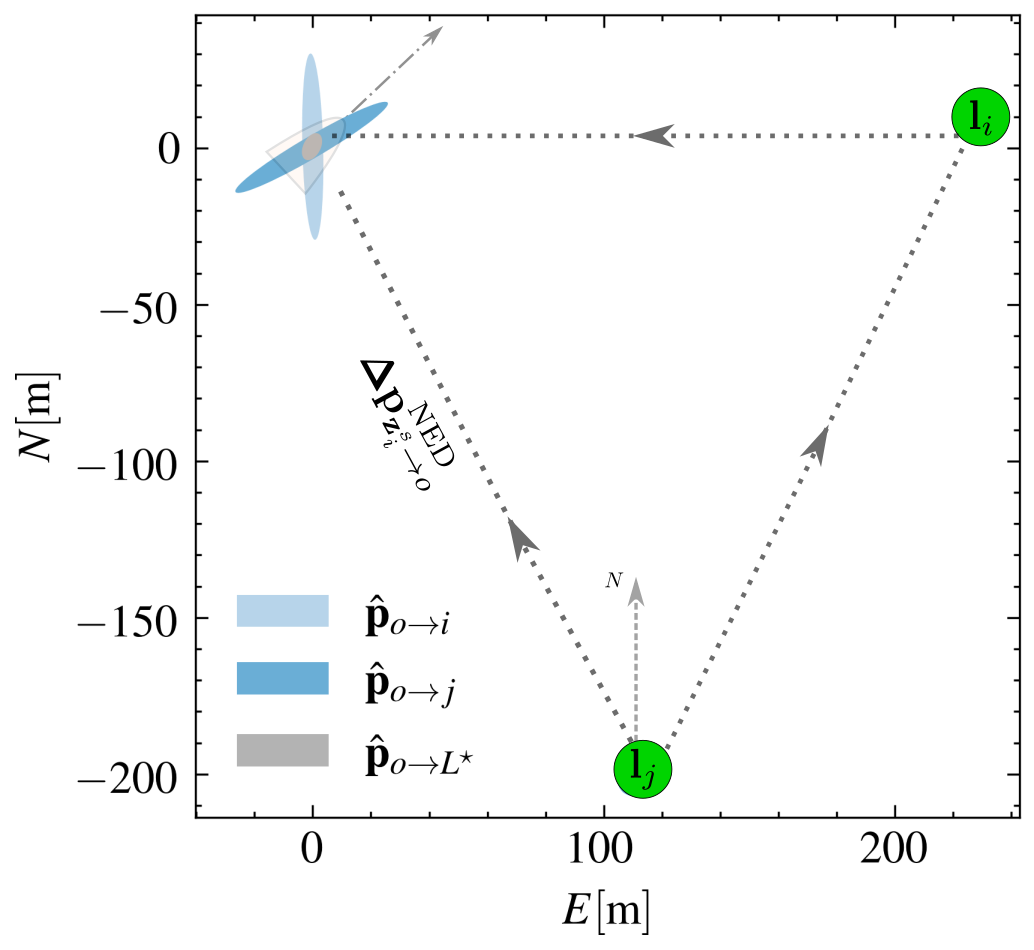}\hspace*{1cm}
		\caption{Resection on a NED plane. The ellipsoids correspond to 95\% confidence intervals, we observe the reduction in uncertainty of the final estimate $\hat{\vec{p}}_{o\to L^{\star}}$ (gray ellipsoid), a product of the probabilistic fusion of the individual position estimates (blue ellipsoids). The estimates are modeled as Gaussian distributions. }
		\label{fig:triangulation_fusion}
	\end{figure}

	In order to move to a metric distance coordinate frame, an arbitrary landmark position $\landmrk_0$ is used as the reference geodetic point, and via the tangent NED plane approximation in \Cref{fig:triangulation}, one can calculate the difference vector $\landmrk_{ij}$ as the vector difference
	\begin{eqnarray}\label{eq:transformation}
	\landmrk_{ij}^{\text{NED}} &=  \landmrk_i^{\text{NED}}   - \landmrk_j^{\text{NED}} \\ &= \mathcal{T}_{\landmrk_0}\left(\landmrk_i , \landmrk_o\right) - \mathcal{T}_{\landmrk_0}\left(\landmrk_j , \landmrk_o\right) \\&= \left(\Delta North, \Delta East\right) =  d_{ij} \phase{a_{ij}}
	\end{eqnarray}
	where
	\begin{itemize}
		\item$\landmrk_{i},\landmrk_{j}  $ are the associated chart landmarks to the static radar tracked targets $\statobservation_{i}, \statobservation_{j} $
		\item$\begin{psmallmatrix}
		\Delta North & \Delta East
		\end{psmallmatrix}^T $ are the North-East components of $\landmrk_{ij}^{\text{NED}}$
		\item$d_{ij}$is the metric distance between the landmarks $i,j$
		\item $a_{ij}$ the angle of the vector $ \landmrk_{ij} $ from \emph{True North} \begin{eqnarray}
		a_{ij} = \arg{<\landmrk_{ij}^{\text{NED}}>}=\arctan{\frac{\Delta \text{East}}{\Delta \text{North}} }   \\
		a_{ji} = \pi + a_{ij}
		\end{eqnarray}
		
		\item $\landmrk_o$ is a reference position in latitude longitude, used as the origin of the local NED coordinates.
	\end{itemize}
	If the origin of the NED system $\landmrk_0$ coincides with the own ship's geodetic location, then the radar observation vectors on the NED plane can be expressed as complex numbers in polar form, and the complex equation relating the edges of the triangle formed between the own ship and the two landmarks is
	\begin{eqnarray}\label{eq:triangle}
	\rho_i e^{j(\hat{\psi}+\beta_i )} -	\rho_j e^{j(\hat{\psi}+\beta_j )}    &= d e^{ja_{ij}}
	\end{eqnarray}
	where $\hat{\psi}$ is the triangulated estimate of the own ship's heading.
	By manipulating \eqref{eq:triangle}
	\begin{eqnarray}
	\rho_i e^{j0} - \rho_j e^{j\beta_{ji}}  = d e^{j(a_{ij}-\beta_i - \hat{\psi})}  \\
	1 - r_{ji}e^{j\beta_{ji}} = \frac{d}{\rho_i} e^{j(a_{ij} - \beta_i - \hat{\psi})}
	\end{eqnarray}
	Hence
	\begin{eqnarray} \label{eq:tanpsi}
	\tan{(a_{ij} - \beta_{i} - \hat{\psi})} = \frac{-r_{ji}\sin{\beta_{ji}}}{1-r_{ji}\cos{\beta_{ji}}}
	\end{eqnarray}
	Then \eqref{eq:tanpsi} can be solved for the own ship's heading estimate $\hat{\psi}$ as
	\begin{eqnarray}\label{eq:hatpsi}
	\hat{\psi} &= \delta \psi + a_{ij} - \beta_{i}
	\end{eqnarray}
	where
	\begin{eqnarray}\label{eq:deltapsi}
	\delta \psi = \arctan{\frac{\sin{\beta_{ji}}\cdot r_{ji}}{1-\cos{\beta_{ji}}\cdot r_{ji}}} = \arctan{\frac{\sin{\beta_{ji}}}{r_{ij}-\cos{\beta_{ji}}}}
	\end{eqnarray}
	Then given the estimated heading $\hat{\psi}$ one can calculate the correction vector of the own ship's pose $\vec{\Delta  p}_{\statobservation_{i}\rightarrow o}^{\text{NED}}$ in polar form
	\begin{eqnarray}\label{eq:deltap}
	\vec{\Delta  p}_{\statobservation \rightarrow o}^{\text{NED}} &= -\rho \phase{\bearing + \hat{\psi}}
	\end{eqnarray}
	with respect to any static target on the radar $\statobservation$. The vector $\vec{\Delta  p}_{\statobservation_{n}\rightarrow o}^{\text{NED}}$ can be converted to geodetic coordinates in order to calculate an estimate of the own ship's geodetic position $\hat{\vec{p}}_{o \rightarrow n}^{\text{geo}}$ by using the associated landmark $\landmrk_{e(n)}$ as reference point to the inverse transformation $\mathcal{T}_{\landmrk_{e(n)}}^{-1} $ as in \eqref{eq:geo_estimate}.
	\begin{eqnarray}\label{eq:geo_estimate}
	\hat{\vec{p}}_{o \rightarrow n}^{\text{geo}}  =\mathcal{T}_{\vec{\landmrk}_n}^{-1}\left(\vec{\Delta  p}_{\statobservation_n \rightarrow o}^{\text{NED}} \right)= \phi(\vec{\kappa}_n) \\
	\vec{\kappa}_n =  \begin{bmatrix}
	\observation_n, \landmrk_n,\hat{\psi}
	\end{bmatrix}, \: n=i,j
	\end{eqnarray}
	\subsection{Probabilistic outputs}
	The above results, are a solution to an over-determined system of equations with three-unknowns (3-DOF pose) and four constraints (range and bearings), whose variables are deterministic.
	
	In accordance to common navigational knowledge, the accuracy of using landmarks for own ship position estimation at sea depends on the angular dispersion between them. The case is slightly different in our presented approach, since we are using both range and bearing information. At the same time, we have to take into account for the various sources of noise in the measurement system. The radar images are sensitive to unwanted noise and interference phenomena (speckle), while at the same time they are being distorted due to the rolling and pitching motion of the vessel on top of which the radar unit is mounted. In practice this means that we have to consider that the extracted landmark features from a radar scan (see again \Cref{fig:buoys_feature_tracking}) are noisy. In that direction, we model the detections as random variables perturbed by additive white noise and in the following section examine the impact of the measurements uncertainty to the estimators accuracy.

	What we are ultimately interested in describing, is the second-stage's pose estimate \eqref{eq:estimateoutput} posterior distribution $p(\hat{\pose}_k|\{\statobservation, \landmrk\}_{i,j} )$, and it's sensitivity to different geometrical landmark configurations at sea.
	
	Now we proceed in defining the p.d.f $ p(\hat{\vec{p}}_{o\to n}^{\text{geo}}) $ through the non-linear transformation \eqref{eq:geo_estimate}
	of the random variables in $\vec{\kappa}_n$. It is assumed that the position of landmark $\landmrk_n$ on the ENC is deterministic and not a random variable.
	
	By a first order linearization of \eqref{eq:geo_estimate} we obtain
	\begin{eqnarray}
	p(\hat{\pose}_k|\{\statobservation, \landmrk\}_{n} )=p(\hat{\vec{p}}_{o\to n}^{\text{geo}}) = \mathcal{N}(\phi(\vec{\kappa}_n), Q_{\kappa_n}) \\
	\phi'(\kappa_n) = \nabla \phi(\vec{\kappa})\rvert_{	\kappa = \kappa_n} \label{eq:first_order1}\\
	Q_{\kappa_n} =		\phi'(\vec{\kappa}_n)\Sigma_{\vec{\kappa}_n} (\phi'(\vec{\kappa}_n))^T \label{eq:first_order2}\\
	\Sigma_{\vec{\kappa}_n} = \begin{bmatrix}
	\var{\range_n} & 0                & 0                \\
	0              & \var{\bearing_n} & 0                \\
	0              & 0                & \var{\hat{\psi}}
	\end{bmatrix}\label{eq:cov_psi}\\
	\kappa_n = \begin{bmatrix}
	\range_n,\beta_n,\hat{\psi}
	\end{bmatrix}^T
	\end{eqnarray}
	As discussed previously, we are interested in defining the posterior pose estimate distribution $p(\hat{\pose}_k|L^{\star} )$. So far, we have only but partially used the amount of information available in $L^{\star}$. Given the two observations and their associated landmarks $\{\statobservation, \landmrk\}_{i,j}$, the own ships heading can be found through \eqref{eq:hatpsi}, but the position can be redundantly estimated using \eqref{eq:geo_estimate} and either associated pairs of observations and landmarks in $\{\statobservation, \landmrk\}_{i,j}\subset L^{\star}$ .
	Since we are interested in using all information available in $L^{\star}$, through the independence assumption and Bayes rule, the final posterior estimate $\hat{\vec{p}}_{o \to L^{\star}}^{\text{geo}}$  that uses all information in $L^{\star}$ is
	\begin{eqnarray}
	p(\hat{\pose}_k|L^{\star} )=p(\hat{\vec{p}}_{o\to L^{\star}}) = p(\hat{\vec{p}}_{o\to i}) p(\hat{\vec{p}}_{o\to j})
	\\=\mathcal{N}(\phi(\vec{\kappa}_i), Q_{\kappa_i}) \mathcal{N}(\phi(\vec{\kappa}_j), Q_{\kappa_j})\\
	=\mathcal{N}\left(\mu_{L^{\star}},\Sigma_{ L^{\star}}\right) \label{eq:final_pos}\\
	\Sigma_{L^{\star}}=\left(\Sigma_{\vec{\kappa}_i}^{-1}+\Sigma_{\vec{\kappa}_j}^{-1}\right)^{-1} \label{eq:fusionSigma}\\
	\mu_{L^{\star}}=\Sigma_{L^{\star}} \left(\Sigma_{\vec{\kappa}_i}^{-1} \phi({\kappa_i})+\Sigma_{\vec{\kappa}_j}^{-1}\phi(\kappa_j)\right) \label{eq:fusionMu}
	\end{eqnarray}
	where \eqref{eq:fusionSigma}\eqref{eq:fusionMu} are a result of the random variables being modeled as multivariate Gaussian distributions.

	Finally, second-stage estimated own ship pose is $\hat{\pose}$
	\begin{eqnarray}\label{eq:estimateoutput}
	\hat{\pose} = \begin{bmatrix}
	\hat{\vec{p}}_{o \to L^{\star}}^{\text{geo}} & \hat{\psi}
	\end{bmatrix}^T
	\end{eqnarray}
	\subsubsection{Heading estimation function}
	We proceed along the lines of the previous section, in order to describe the distribution of the estimated heading $\hat{\psi}$. We rewrite
	\eqref{eq:deltapsi} as the non-linear transformation
	\begin{eqnarray}\label{eq:nonlineartransformation}
	\delta\psi = f(\beta_{ji}, r_{ji})
	\end{eqnarray}
	Where $\beta_{ji}$ and $r_{ji}$ are now random variables describing the bearing difference and range ratio  respectively. Thereafter, we seek to determine the probability density functions of the range ratio $p(r_{ji})$ and bearing difference $p(\bearing_{ji})$.
	\subsubsection{Random variables and approximations}
	Under the assumption that the range and bearing measurements, are perturbed by additive, white Gaussian noise
	\begin{eqnarray}
	p(\range) &= \gaussian{\mu_{\range}}{\var{\range}} \\
	p(\bearing) &= \gaussian{\mu_{\bearing}}{\var{\bearing}} \label{eq:randombearing}
	\label{eq:randomvariables}
	\end{eqnarray}
	It follows from  \eqref{eq:rangeratio}\eqref{eq:bearingdiff} that
	\begin{eqnarray}
	p(r_{ji}) &= \frac{\gaussian{\mu_{\range_j}}{\var{\range_j}}}{\gaussian{\mu_{\range_i}}{\var{\range_i}}}\label{eq:randomvariables1}\\
	p(\bearing_{ji}) &= \gaussian{\mu_{\bearing_j} - \mu_{\bearing_i}}{\var{\bearing_j} + \var{\bearing_i}-2\var{\bearing_j\bearing_i}}
	\label{eq:randomvariables2}
	\end{eqnarray}
	In reality, two targets  $\observation_{i,j}$ are at least partially correlated, owing to the fact that the tracking module is using the same radar unit to detect them.  we assume independence for the shake of simplification.
	\begin{eqnarray}\label{eq:crosscovariances}
	\var{\bearing_j\bearing_i}&= \var{\range_j\range_i} &\approx 0 ,\quad i\neq j
	\end{eqnarray}
	While fully closed form expressions for the ratio of non-central, Gaussian variables $p(r_{ij})$ exist (\cite{Diaz-Frances2013}), we find them cumbersome and impractical, thus we approximate the parametric p.d.f. of the ratio of the two non-central Gaussian distributions in \eqref{eq:randomvariables1} with a normal distribution (\cite{Diaz-Frances2013})as in \eqref{eq:ratioapprox}, under the conditions in \eqref{eq:ratioapproxconditions}.
	\begin{eqnarray}
	p(r_{ji}) \simeq \gaussian{\frac{\mu_{\range_j}}{\mu_{\range_i}}}{\frac{\mu_{\range_j}^2}{\mu_{\range_i}^2}\bigg(\frac{\var{\range_i}}{\mu_{\range_i}^2}+\frac{\var{\range_j}}{\mu_{\range_j}^2}\bigg)}
	\label{eq:ratioapprox}
	\end{eqnarray}
	\begin{eqnarray}
	\frac{\sigma_{\range_i}}{\mu_{\range_i}}<1
	\,		\land \, 		\frac{\sigma_{\range_j}}{\mu_{\range_j}}<1 \, \land \, \sigma_{\range_i\range_j}\simeq 0
	\label{eq:ratioapproxconditions}
	\end{eqnarray}
	These conditions are very valid for a radar unit and targets at sea where it is common for the true value of a target's range to be much larger than the standard deviation of the noise in the range measurements.

	\subsubsection{Second order Taylor approximation of  $f$ }
	In order to obtain a parametric description for the probability density function $p(\delta\psi)$ in \eqref{eq:deltapsi} , we introduce a a 2nd-order Taylor approximation (\cite{Hendeby2003}) on \eqref{eq:nonlineartransformation}. Then the distribution $p(\delta\psi)$ can be approximated with a Gaussian \eqref{eq:gaussapprox} with parameters expressed as functions of the input parameters $\boldsymbol{\mu_x} , \Sigma_x$ \eqref{eq:mux}\eqref{eq:sigmax}
	%
	
	%
	\begin{eqnarray}
	\delta\psi &= f(x),\ x = \begin{pmatrix}\beta_{ji}, r_{ji}\end{pmatrix}^T \label{eq:nonlinearpsi}\\
	p(\delta\psi) &\simeq \gaussian{\mu_{\delta\psi}}{\var{\delta\psi}} \label{eq:gaussapprox}  \label{eq:pdfdeltapsi}\\
	\mu_{\delta\psi} &= f(\boldsymbol{\mu_x}) + \tfrac{1}{2} \tr\left(f''(\boldsymbol{\mu_x})\Sigma_x\right) \\
	\var{\delta\psi} &= f'(\boldsymbol{\mu_x})\Sigma_x(f'(\boldsymbol{\mu_x}))^T \\
	&+ \tfrac{1}{2}\tr\left(\Sigma_x f''(\boldsymbol{\mu_x})\Sigma_x f''(\boldsymbol{\mu_x})\right)
	\end{eqnarray}
	Where $f'$ and $f''$ are the Jacobian and the Hessian matrices respectively, evaluated at $\boldsymbol{\mu_x}$
	\begin{eqnarray}
	\boldsymbol{\mu_x} = \begin{pmatrix}
	\mu_{r_{ji}},\mu_{\bearing_{ji}}
	\end{pmatrix}^T \label{eq:mux}\\
	\Sigma_x = \begin{bmatrix}
	\var{r_{ji}} & 0 \\ 0 & \var{\bearing_{ji}}
	\end{bmatrix} \label{eq:sigmax}\\
	f'(\boldsymbol{\mu_x}) = \nabla f(x)\rvert_{x = \boldsymbol{\mu_x}} \label{eq:fprime}\\
	f''(\boldsymbol{\mu_x}) = \mathbf{J} \left(\nabla f(x)\right) \rvert_{x = \boldsymbol{\mu_x}} \label{eq:fdoubleprime}
	\end{eqnarray}
	Thereafter, from \eqref{eq:hatpsi} and the random variable properties \eqref{eq:randombearing}\eqref{eq:pdfdeltapsi}, the p.d.f. of the estimated triangulated heading is
	\begin{eqnarray}
	p(\hat{\psi}) &= \gaussian{\mu_{\delta\psi}+a_{ij}-\mu_{\bearing_i}}{\var{\delta\psi}+\var{\bearing_i}}\label{eq:pdfheading}
	\end{eqnarray}
	
\begin{figure*}
		\centering
		\includegraphics[width=\textwidth]{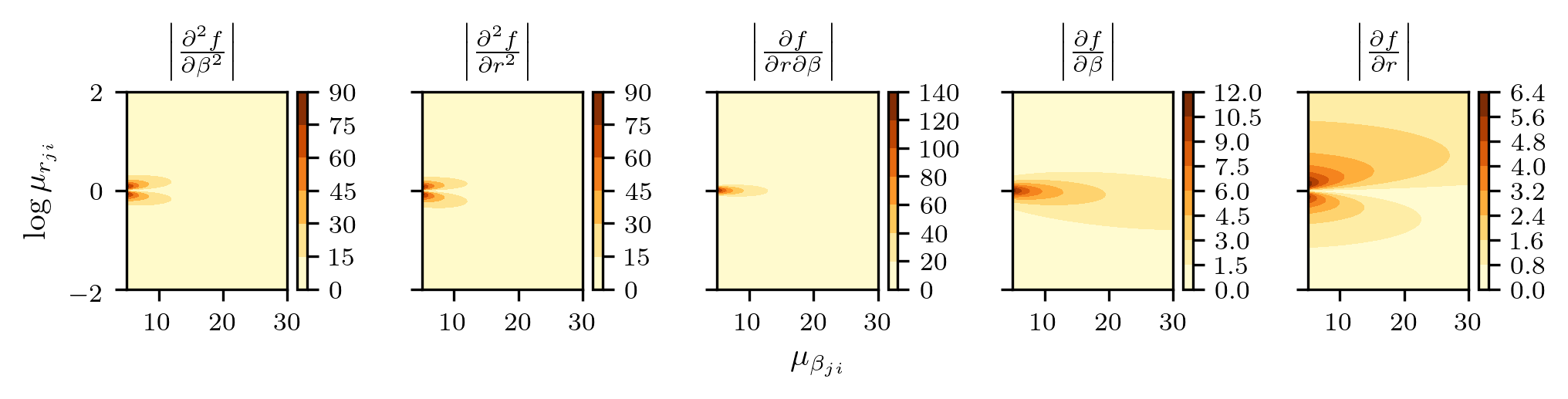}
		\caption{Elements of the gradient $\nabla f(
			\mu_{r_{ji}},\mu_{\bearing_{ji}}
			)$\eqref{eq:fprime} and the Hessian  $\mathbf{H} \left(f(	\mu_{r_{ji}},\mu_{\bearing_{ji}}) \right)$ \eqref{eq:fdoubleprime}}
		\label{approximation_quality}
	\end{figure*}

	The variance of the heading estimation $\var{\hat{\heading}}$ subsequently affects the position estimation in \eqref{eq:deltap}, through the covariance matrix in \eqref{eq:cov_psi}. We therefore investigate geometric setups that could lead to high values of $\var{\hat{\heading}}$. These particular geometric setups can be exposed from the contour plots of the elements of the gradient and the Hessian (\Cref{approximation_quality}) in \eqref{eq:fprime}\eqref{eq:fdoubleprime}.
	. As a general observation,  it is noted that the closer the landmarks get to each other,  $\log \mu_{r_{ji}}\approx \mu_{\bearing_{ji}} \approx 0$, the larger the magnitude of the elements in the gradient and Hessian matrices , therefore the larger the impact of the measurement's uncertainty to the variance of the estimated heading ${\var{\heading}}$ as propagated through \eqref{eq:nonlinearpsi}.

	\begin{figure}[ptb]
		\centering
		\includegraphics[width=1\columnwidth]{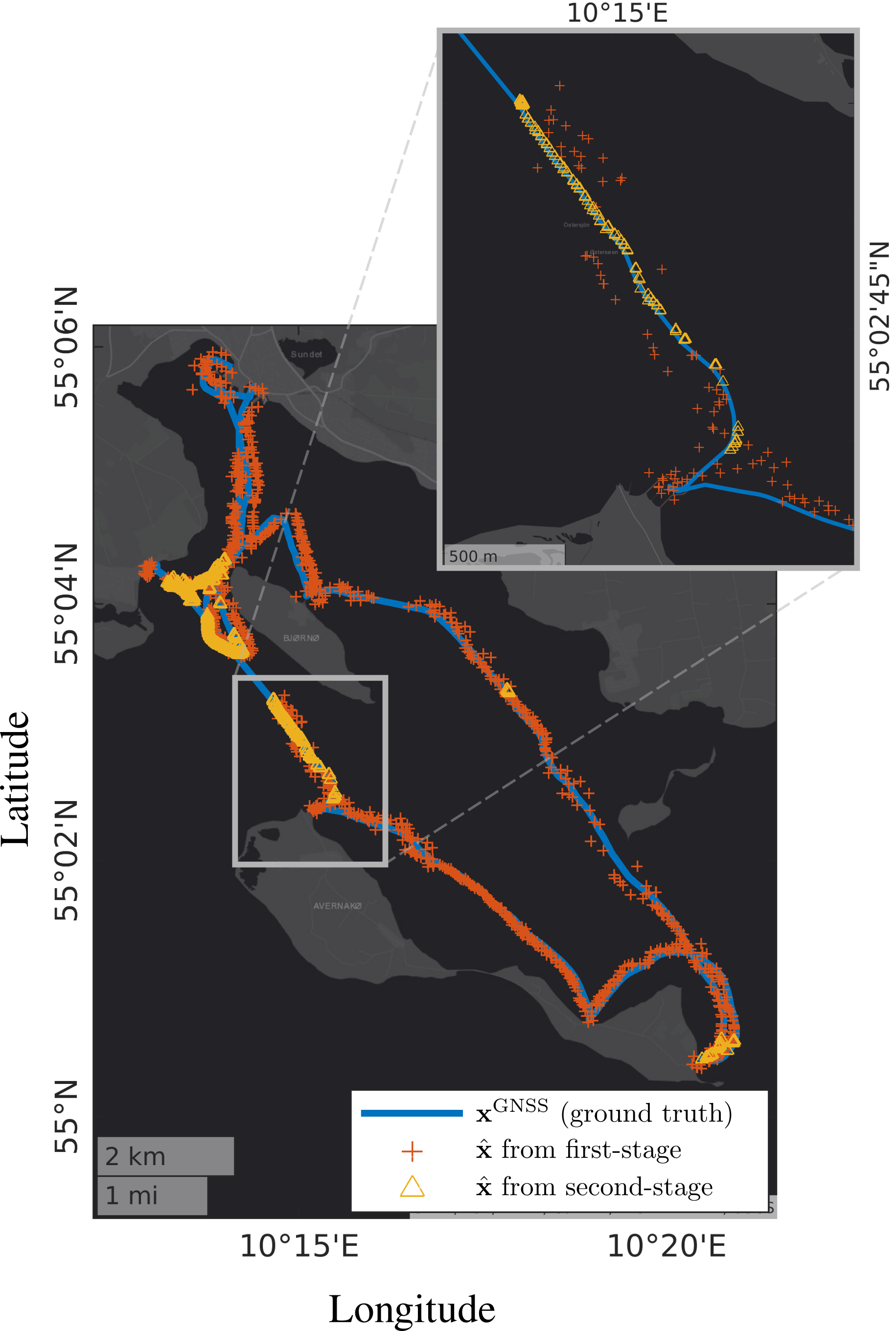}
		\caption{Performance of the two estimators on data captured at sea. In orange the result of the shoreline-matching estimator, and in yellow the result of the second-stage estimation. It is evident that buoys are not always available to perform resection with, but the estimated pose is much more accurate than the pose from the likelihood-field estimator. Data captured on March 2021.}
		\label{fig:two_stage_output}
	\end{figure}
	
	\subsection{Position estimation summary}
\begin{figure}[ptb]
		\centering
		\includegraphics[width=1.0\columnwidth]{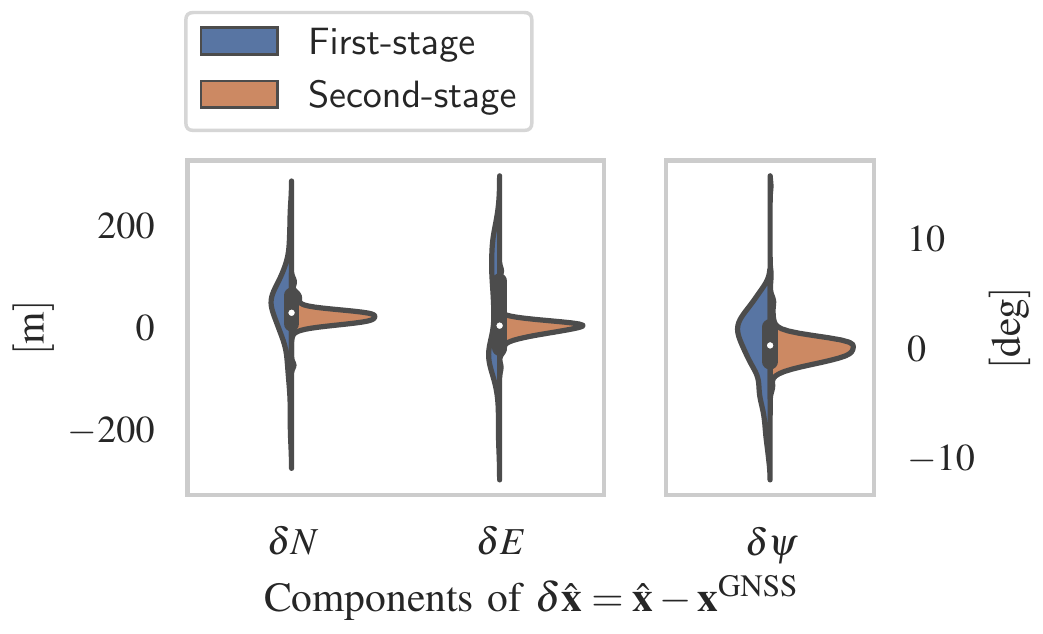}
		\caption{Error distributions between the estimated pose from the individual stages and the GNSS measurements as ground-truth $\delta\hat{\mathbf{\pose}}=\hat{\mathbf{x}}-\mathbf{x}^{\text{GNSS}}$. The second-stage estimates are much more accurate, but not always available.}
		\label{fig:error_distributions}
	\end{figure}

	So far, we have described an independent positioning system from the GNSS, relying on a two-stage pose estimation process in which
	\begin{enumerate}[label=(\alph*)]
		\item The first stage uses shoreline features and matches them with similar features on the ENC, deriving a pose estimate $\poseest_k$. The estimation is consistent but it's accuracy varies depending on the morphology of the shoreline and the map's accuracy. 
		\item The second stage detects nearby buoys or beacons on the radar, associates them with same objects on the ENC and derives the own ship's position through a combined triangulation-trilateration approach. This process results in an estimate $\hat{\pose}_l$ of refined accuracy compared to the first-stage. The difference in accuracy between the two stages can be seen in \Cref{fig:two_stage_output,fig:error_distributions}. An important limitation of the second-stage estimator, is it's limited operational domain. More in specific, not all sailing paths are traversed around buoys, and not all buoys are equipped with radar reflectors. This limits the availability to perform the second-stage estimation to specific sailing areas as seen in \Cref{fig:two_stage_output}. Despite its higher accuracy, its limited availability encourages us to avoid using it in the design of the monitoring system following in the next sections.
	\end{enumerate}
	\Cref{tab:estimator_table} summarize strength's and weaknesses of each estimation stage, while \Cref{fig:error_distributions,fig:two_stage_output} illustrate the difference in accuracy and availability of each estimate.  
	
	\begin{table}[]
		\centering
		\caption{Comparing outputs of different stages of the estimation process. Error calculated as  $\delta\pose = \hat{\pose}-x^{\text{GNSS}}$}
		\renewcommand{\arraystretch}{1.1}
		\begin{tabular}{m{0.2\columnwidth}m{0.18\columnwidth}m{0.45\columnwidth}}
			\toprule
			$\hat{\vec{x}}_k$& First-stage  & Second-stage  \\ \midrule
			Error STD ${\delta\hat{\vec{\pose}}}$    & \SI{61}{\meter}  & \SI{21}{\meter}   \\
			Availability    & Always & Only in sailing regions that include beacons-buoys with radar reflectors \\
			Robustness & High & Depends on the first-stage estimate to solve the data association\\
			\bottomrule
		\end{tabular}
		
		\label{tab:estimator_table}
	\end{table}
	
	\subsection{Statistical change detector}
	In view of an imminent cyber-attack or failure on the GNSS receiver, we wish to be able to detect and mitigate such events and provide alarms to the navigator. The independent positioning system discussed in the previous sections, allows us to monitor the integrity of the GNSS information. We are doing so by statistically detecting changes in the nominal  behaviour of the estimates. In the design of the detector, we selectively utilize the position estimates as delivered by the first-stage only, disregarding the second-stage. The motivation behind this decision, is that we wish to monitor under all operating conditions, irregardless of the presence of nearby beacons or buoys, whose scarcity in a regular mission at sea can be high, as illustrated in \Cref{fig:performance_comparison_geo}.
	
	\subsubsection{Nomenclature}
	To describe the monitoring system, we adopt the following nomenclature
	\begin{itemize}
		\item[] $\posegps\ofk$ $ \in \mathbb{R}^2$ Position information as delivered through  satellite GNSS compass, namely latitude, longitude.
		\item[] $\pose^R\ofk \in \mathbb{R}^2$ Estimated position, latitude and longitude, as delivered through the first stage estimation, or the second-stage is available.
		\item[] $k \in \mathbb{Z}^+$  denotes the sampling index.
		\item[] $t\ofk$ denotes the sample time $k$.
	\end{itemize}
	\subsubsection{Residual}
	We fuse information across different sensors, by defining the residual quantity $r\ofk \in \mathbb{R}^{+}$ as the distance error between the reported position from the GPS and the estimated position from first-stage estimator, when transformed into an NED frame.
	\begin{eqnarray}\label{eq:residual}
	r\ofk =\|\mathcal{T}_{\posegps\ofk} (  \posegps\ofk \ -  \pose^R\ofk )\|
	\end{eqnarray}
	The residual in \eqref{eq:residual} is a by-product of a multi-modal fusion of information, where from a diagnostic point of view, we perform \emph{statistical change detection} on $r_k$, in order to detect deviations from nominal behaviors.
	
	The information delivered by the GNSS and the Radar is subjected to random noise, so $r\ofk$ is a sample of the random variable $R$ with probability density function $p_{R}(x)$
	\begin{eqnarray}
	r\ofk \sim p_{R}(x)
	\end{eqnarray}
	A certain description for the p.d.f. $p_{R}(x;\boldsymbol\theta)$ arises if we assume a parametric description according to a known family of distributions, where $\mathbf{\boldsymbol\theta}$ corresponds to the vector of parameters.
	\subsubsection{Hypothesis testing}
	Under a parametric description for $p_{R}(x)$, the following hypotheses are formed, as means of describing the different conditions of the system.
	\begin{eqnarray}
	r\ofk \sim p_R(x;\params)\\
	\mathcal{H}_0 : \params = \params_0 \\
	\mathcal{H}_1 : \params \neq \params_0
	\end{eqnarray}
	Under a compact parametric description, where the underlying distribution is described by a Gaussian p.d.f, we can summarize all information about $\hy_0$ in the parameter vector $\params_0 \in \mathbb{R}^2$, namely the mean and the variance of the distribution and describe the nominal conditions by the parameter vector $\params_0$.
	
	While in many applications this is a sound approach, in many other cases, defining a unique $\params_0$ vector does not suffice to accurately describe the residual's properties, more so because, the residual being practically only but a realization of a physical random process, whose stationarity can not be ensured, and thus the parameter vector depends on $t_k$.
	\subsubsection{General Likelihood Ratio Test (GLRT)}
	The residual time-series $\res=\{r_0, r_1, r_2,...,r\ofk\}$  is distributed according to $p_R(r;\params_0; \hy_0)$ under $\hy_0$, and according to $p_R(r;\params_1; \hy_1)$ under $\hy_1$. As already specified, the forms of the p.d.f.s as well as the dimensonalities of the unknown parameter vectors $\params_0,\params_1$ can vary under each hypothesis. The test statistic $g_k$ is
	\begin{eqnarray}\label{eq:test_statistic}
	g_k =\log L_G(\vec{r}) = \log \frac{p_R(\vec{r};\hat{\params}_1;\hy_1)}{p_R(\vec{r};\hat{\params}_0;\hy_0)}
	\end{eqnarray}
	where
	\begin{eqnarray}
	\hat{\params}_i = \displaystyle \underset {\params_i }{\operatorname {arg\;max} }\,p_R(\vec{r};\hat{\params}_i;\hy_i)
	\end{eqnarray}
	are the \emph{Maximum Likelihood Estimates} of $\params_i$ under each hypothesis.
	
	Then the GLRT decides $\hy_1$ or $\hy_0$ depending on the threshold $\gamma$
	\begin{eqnarray}
	g_k \overset{\overset{\mathcal{H}_1}{>}}{\underset{\underset{\mathcal{H}_0}{<}}{}}\gamma
	\end{eqnarray}
	\subsubsection{Parametric vs Non-parametric}
	It is a common assumption in change detection to consider that the time-series
	\begin{eqnarray}
	\vec{r}=\{r_0, r_1, r_2,...,r_k\}
	\end{eqnarray}
	comprises of \emph{independent and identically distributed} samples drawn from the same unknown density $p_R(r)$. It is common for $p_R(r)$ to assume a parametric model from the family of exponential distributions, more in specific the Gaussian model. While convenient and powerful, the Gaussian distribution is uni-modal and in several instances not flexible enough to capture the statistical structure in the  data $\vec{r}$ (see \Cref{fig:models}.
	
	Motivated by the flexibility of non-parametric distributions, we hereby propose a Kernel Density Estimate (KDE) for the p.d.f. $p_R(r)$.
	We are interested in estimating the shape of $p_R(r)$, hence it's KDE is
	\begin{eqnarray}
	{\displaystyle {\widehat {p}}_{R}(r;\vec{r})={\frac {1}{nh}}\sum _{i=1}^{n}\Phi{\Big (}{\frac {r-r_{i}}{h}}{\Big )},\ r_i \in \vec{r}}
	\label{eq:kde}
	\end{eqnarray}
	
	where
	
	\begin{itemize}
		\item $\Phi(x)$ is the Kernel function, ${\displaystyle \Phi (x)={\frac {e^{-{\frac {x^{2}}{2}}}}{\sqrt {2\pi }}}}$, a non-negative function, in our case chosen to be the standard normal density function, due to it's convenient mathematical properties.
		\item $h$ is the bandwidth parameter, and is chosen based on cross-validation of the detector's performance.
	\end{itemize}
	Therefore we define the test statistics $g_k^{\text{Gauss.}}$ and $g_k^{\text{KDE}}$, depending on the model used to approximate $p_R(\vec{r})$ in \eqref{eq:test_statistic}.
	\comment{
		\subsubsection{Bandwidth}
		The bandwidth selection greatly influences the performance of the detector. As the scaling parameter $h$ trends to the two following extrema, we observe different behaviors (Figure \ref{fig:kde_h}), namely
		
		\begin{itemize}
			\item As $h\to 0$, the locality of the kernel is exaggerated, to the point that there is no smoothing, and the estimate is a sum of delta functions centered at the coordinates of the analyzed samples $x_i$.
			\item As $h \to \infty$, locality diminishes, and the estimate is returning the actual kernel shape, but centered on the mean of the sample population.
		\end{itemize}
		
		\begin{figure}
			\centering
			\caption{Kernel Density Estimation of a random sample of 100 points, drawn from the standard normal distribution, for varying bandwidth values. In grey: the true density. In red: KDE with h=0.05. Black: KDE with h=0.337. Green: KDE with h=2. (By User:Drleft, CC BY-SA 3.0, https://commons.wikimedia.org/w/index.php?curid=73892711)}
			\label{fig:kde_h}
		\end{figure}
	}

	\subsubsection{Quasi-stationary residuals}
\begin{figure}
		\centering
		\includegraphics[width=1\columnwidth]{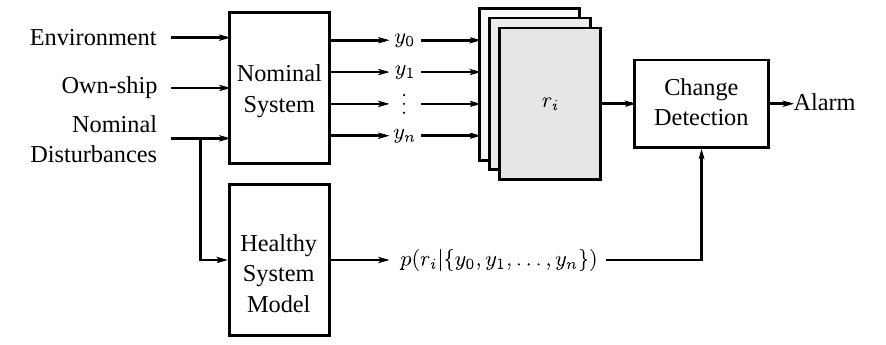}
		
		\caption{Non-adaptive model-based statistical change detection. Variations in the Nominal disturbances or in the system's model result in quasi-stationary residuals and give rise to false alarms. }\label{fig:fd}
	\end{figure}

	A classical example of a change detection architecture is illusrated in \Cref{fig:fd}. Observable variables $y_{i}$ are either directly measured or estimated, $r_{i}$ are residuals generated from the structural analysis of the healthy system model. The statistical change detection module is providing alarms based on deviations of the residual's characteristics from the nominal behavior. These healthy system is described by parameters found after fitting a general model on data captured under presumably nominal conditions. Detection capabilities of such an architecture are great, depending on the model's accuracy, the validity of the underlying assumptions, and the fact that the system's nominal behavior, as well as that of the external disturbances, do not evolve over time. This is an important limitation, allowing temporal changes in the model or in the nominal disturbances to give rise to false alarms. In our example, it was observed that the residual's expected value and distribution vary over time, depending on the morphology of the shore-line and the accuracy of the maps. A shore-line is by nature a seasonal and long-term varying feature (\cite{Stive2002}\cite{Pianca2015}), especially in shallow-water areas such as Denmark. Additionally, the accuracy of the map depends on how recently it has been designed, and how sensitive the mapped area is to shoreline changes. Map and terrain variability directly influence the behavior of the residual $\vec{r}$ that we intent to perform monitoring on and add make it quasi-stationary under nominal conditions. This is illustrated in \Cref{fig:panel}, where during the first 20 minutes of the mission, we observe a reduction in the variance of $r(k)$, which suddenly increases around $t=\SI{35}{\min}$. Directly emerges the need thus, for the system to adapt to the changes in the residual's behavior.
	\subsubsection{Double window approach}
	By adopting a double moving window strategy over the residual time-series $\vec{r}$, the test-statistic is adapting to long-term variations of the residual, while still remaining sensitive to short-term variations such as a spoofing incident.
	
	Two moving windows of length $\lambda$ and $\mu$, separated by a gap of length $o$ are being slided over the residual time-series $\vec{r}$, and digesting continuous chunks of data from it. This is described by the following moving subsets
	\begin{eqnarray}
	\vec{r}^L_k = \{r(k-\lambda), r(k-\lambda +1),...,r\ofk\} \\
	\vec{r}^M_k = \{r(k-\mu- o), r(k-\mu - o +1)\\,...,r(k-o)\}
	\end{eqnarray}
	where $\lambda,\mu, o$ are the window lengths of $L,M$ and the gap in samples between the $L$ and the $M$ window respectively.
	
	In principle, since $\vec{r} \sim p_R(r;\hy_0)$, with the elements of $\vec{r}$ being independent and identically distributed, then the subsets $\vec{r}^L_k,\vec{r}^M_k $ should be distributed according to the same distribution.
	\subsubsection{Composite hypothesis}
	This allows us to define the composite null hypothesis $\hy_0^\star$
	\begin{eqnarray}\hy_0^\star:\begin{array}{l}
	\vec{r}^L_k \sim  p_R(r;\hy_0) \\
	\vec{r}^M_k \sim  p_R(r;\hy_0)
	\end{array}
	\end{eqnarray}
	With the double moving window approach, we expect that under the occurrence of a change, the residual inside the two windows will be distributed among different hypothesis, thus we can define  $\hy_1^\star$
	\begin{eqnarray}\hy_1^\star:\begin{array}{l}
	\vec{r}^L_k \sim  p_R(r;\hy_0) \\
	\vec{r}^M_k \sim  p_R(r;\hy_1)
	\end{array}
	\end{eqnarray}
	In other words:
	\begin{itemize}
		\item[$\hy_0^\star$] implies that both $\vec{r}^M_k$ and $\vec{r}^L_k$ describe the same underlying distribution.
		\item[$\hy_1^\star$] implies that $\vec{r}^M_k$ is describing a different distribution than the one explained by $\vec{r}^L_k$.
	\end{itemize}
	Then from \eqref{eq:kde} the GLR test becomes
	\begin{eqnarray}
	L^{\text{KDE}}\ofk = \frac{p_R(\vec{r}; \hy^\star_1)}{p_R(\vec{r};\hy^\star_0)}=\displaystyle\prod_{r \in \vec{r}^M_k}\frac{{\widehat {p}}_{R}(r;\vec{r}^M_k)}{{\widehat {p}}_{R}(r;\vec{r}^L_k)}
	\end{eqnarray}
	Then taking the log-likelihood, the GLR test decides $\hy^\star_1$ if
	\begin{eqnarray}
	g^{\text{KDE}}\ofk = \log \displaystyle\prod_{r \in \vec{r}^M_k}\frac{{\widehat {p}}_{R}(r;\vec{r}^M_k)}{{\widehat {p}}_{R}(r;\vec{x}^L_k)} > \gamma
	\end{eqnarray}
	or
	\begin{eqnarray}
	g_k \overset{\overset{\hy^\star_1}{>}}{\underset{\underset{\hy^\star_0}{<}}{}}\gamma
	\end{eqnarray}

	\subsubsection{Detectors in parallel}
	
	We have so far presented two detectors, one based on a Gaussian GLRT and one based on a KDE GLRT. We have also discussed how the performance of each detector compares to the other depending on the validity of the assumptions governing the residual time-series $\vec{r}$.
	
\begin{figure}
		\centering
		\includegraphics[width=\columnwidth]{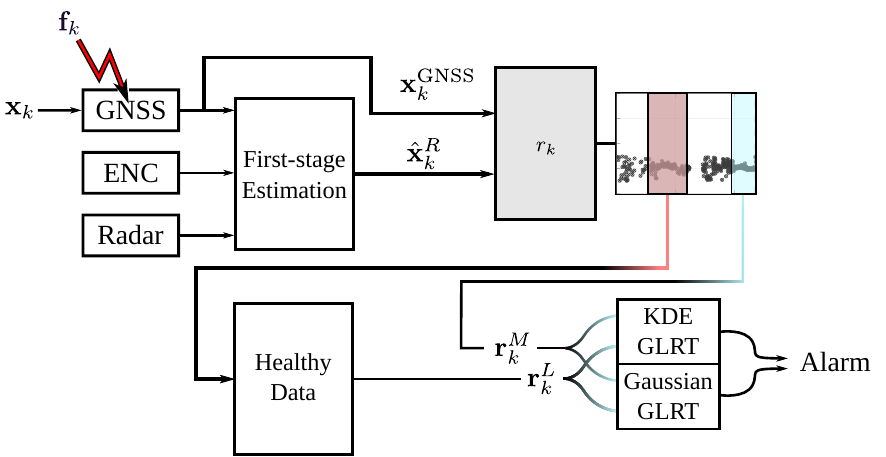}
		\caption{Combined detector approach. The healthy system's model is replaced with an adaptive statistical representation, fitted to recent data $\vec{r}^{L}_k$.}
		\label{fig:double_windowBlock}
	\end{figure}

	Within a parametric Gaussian representation, the model is represented by the common exponential form and its moments. It is known that the Gaussian GLRT is the most powerful test, given that the assumptions about the underlying data are true. In the KDE representation, the data \emph{are} the model. In that sense the KDE is a more flexible representation, and thus describes the residual distribution more accurately in the complementary cases where statistical structure in the residual can not be sufficiently captured by the simple exponential form. This complementary nature of the detectors, leads naturally to us using them in parallel (see \Cref{fig:double_windowBlock}).
	
	Given the test-statistic outcome of the two detectors $g^{\text{Gauss.}}_k , g^{\text{KDE}}_k$ and their associated thresholds $\gamma_{\text{Gauss.}},  \gamma_{\text{KDE}}$
	
	Then the combined GLRT raises an alarm if
	
	\begin{eqnarray}
	(g^{\text{Gauss.}}_k > \gamma_{\text{Gauss.}}) \lor (g^{\text{KDE}}_k > \gamma_{\text{KDE}} )
	\end{eqnarray}

	\subsection{Reasoning}
\begin{figure}[ptb]
		\centering
		\includegraphics[width=1\columnwidth]{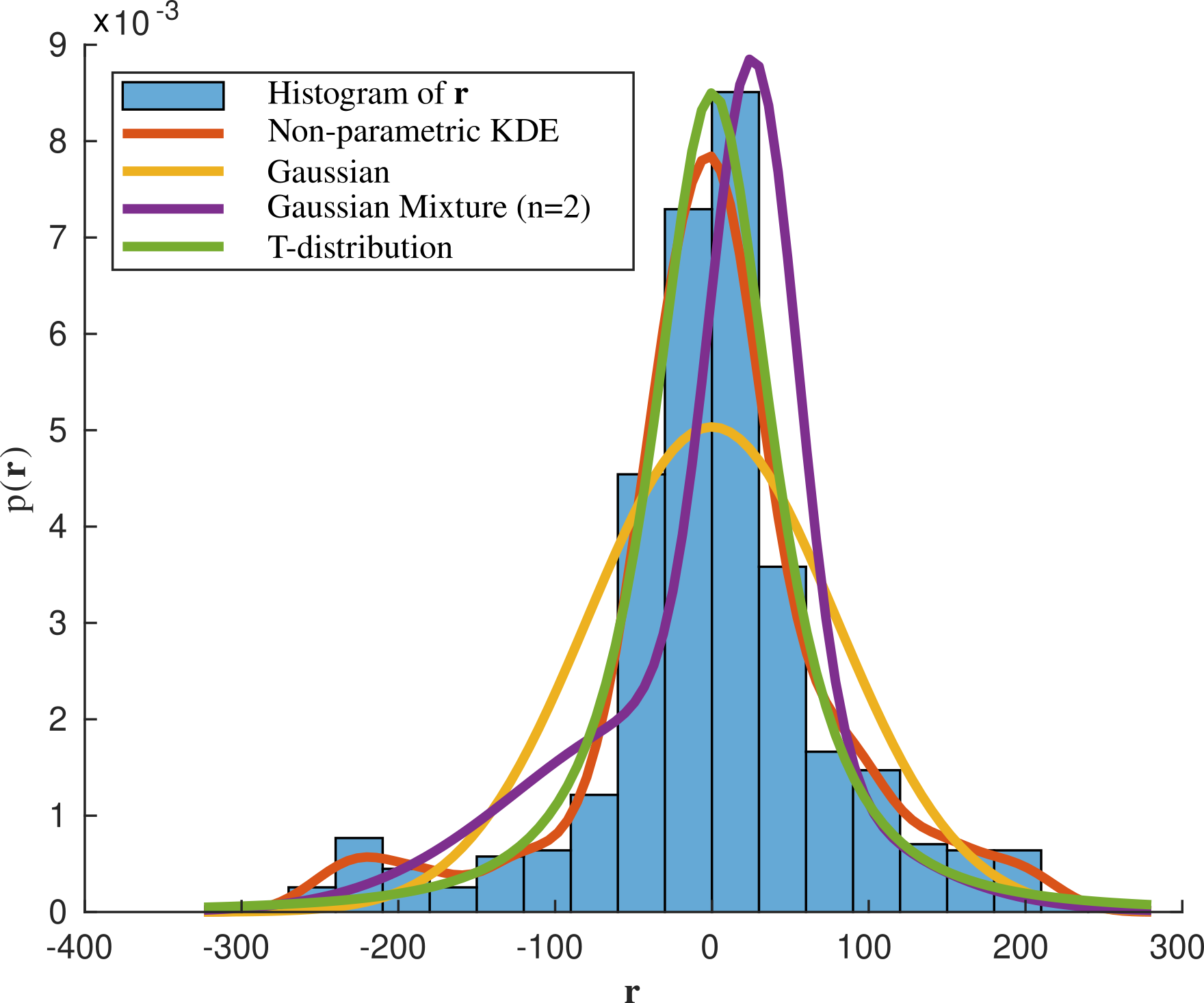}
		\caption{Empirical distribution of the residual $\vec{r}$ under nominal conditions $\vec{r} \sim p_R(r;\hy_0)$, and fitted distribution models. One can observe how the flexibility of the KDE-model (red) allows the estimate to fit on the underlying normalized histogram (blue), better than the parametric models, which fail to capture the bi-modalities that appear near the tails.}
		\label{fig:models}
	\end{figure}

	General parametric descriptions, albeit powerful, are inflexible and limit the detection capabilities when the modelling and independence assumptions are violated, as commonly seen on real-life deployed systems (see Figure \ref{fig:models}). The aforementioned, KDE, double-window approach is learning on-line the statistical properties of the residual $r\ofk$ and is detecting short-term changes in it, where short-term is defined by the window sizes and the gap between them.

	Summarized, the main ideas in the presented approach are:
	\begin{enumerate}[label=(\alph*)]
		\item A \emph{double window} approach ensures adaptability in long-term variations of the residual properties.
		\item \emph{Kernel Density Estimation} allows the required amount of flexibility needed when dealing with distributions that deviate from known parametric forms and improves detection times.
	\end{enumerate}

	In the parametric Gaussian representation, the model is represented by the common exponential form and its moments. It is known that the Gaussian GLRT is the most powerful test, given that the assumptions about the underlying data are true. In the KDE representation, the data \emph{are} the model. In that sense the KDE is a more flexible representation, and thus describes the data more accurately, in the complementary cases where statistical structure in the data can not be sufficiently captured by the simple exponential form.

	\subsubsection{Parameters}
	
	The detector is characterized by multiple tuning parameters that need to be identified,
	
	\begin{itemize}
		\item KDE GLRT
		\begin{enumerate}[label=(\alph*)]
			\item Window lengths $\mu,\lambda,o$
			\item Bandwidth $h$
			\item A threshold $\gamma_{\text{KDE}}$
		\end{enumerate}
		\item Gaussian GLRT
		\begin{enumerate}[label=(\alph*)]
			\item Similarly, to the KDE GLRT, window lengths $\mu,\lambda,o$
			\item A threshold $\gamma_{\text{Gauss.}}$
		\end{enumerate}
	\end{itemize}
	
	Selection of optimally performing parameters is discussed in the Results sections.
	\subsection{Threshold selection}
	
	The threshold selection is defined by the probability of false alarm $P_{FA}$, and the Empirical Cumulative Distribution function $F_g$ of the test-statistic $g\ofk$ under nominal system conditions, that is when we know that the GNSS is reliable.
	
	\begin{eqnarray}
	\gamma  =\,\min \left\{x\in {\mathbb  {R}}:1-P_{FA}\leq F_{g}(x)\right\}
	\end{eqnarray}

	The imposed design constraint is the mean time between false alarms $t_{FA}$. We  impose $t_{FA}= 1 \text{ year}$ as a requirement, then given a mean time between samples $\bar{T}_s = 4.96 \text{ s}$
	\begin{eqnarray}
	P_{FA} &= \frac{T_s}{t_{FA}} = 1.571763\times 10^{-7}
	\end{eqnarray}
	Depending on the type of model used for the test statitic $g\ofk$, we derive individual thresholds $\gamma_{\text{Gauss.}},  \gamma_{\text{KDE}}$ corresponding to the test-statistics $g^{\text{Gauss.}}_k , g^{\text{KDE}}_k$.
	\section{Results}
	We speculate on a spoofing incident similar to the one in the work of \cite{Bhatti2017a}, where
	the nominal GNSS position is being disturbed by an additive cross-track error $f\ofk$ (see \Cref{fig:double_windowBlock}
	\begin{eqnarray}
	f\ofk=f_{\text{slope}}\cdot (t\ofk-t_{\text{onset}})
	\end{eqnarray}
	The residual $r\ofk$ is generated from \eqref{eq:residual}. The panel in Figure \ref{fig:panel} compares the performance of the non-parametric KDE GLRT, to that of a Gaussian GLRT.
	\begin{itemize}
		\item [$t_{\text{onset}}$] signifies the moment that the fault begins to occur.
		\item [$f_{\text{slope}}$] is the rate of change of the additive cross-track fault, in this case $20 \frac{\si{m}}{\si{min}}$.
		\item [$t_D$] is the time to detect the change, or else the time that has elapsed since  $t_{\text{onset}}$ at the moment of detection.
	\end{itemize}
	\subsubsection{Detector parameters}
	The optimally performing parameter sets are found by maximizing the individual detector's performance. In order to do so, we create different spoofing attack realizations, in which the onset-time spans across the mission. This is performed for 1000 different cases $i$, where $t^i_{\text{onset}}$ is uniformly distributed across the span of the mission's duration. The time to detect $t^i_{D}$ is measured for case $i$ and we are minimizing the performance index $J = \sum_{i=1}^{1000} {t^i_{D}}^2$  over the detector parameters performance index. The process is the same among both detectors, with the inclusion of the extra bandwidth parameter $h$ for the KDE GLRT. The optimal parameter set for each detector is thus
	\begin{eqnarray}
	\begin{pmatrix} \mu, & \lambda, &o, &h \end{pmatrix}_{\text{KDE}} = \displaystyle \underset {\begin{psmallmatrix} \mu, & \lambda, &o, &h \end{psmallmatrix}}{\operatorname {arg\;min}  J_{\text{KDE}}}\\
	\begin{pmatrix} \mu, & \lambda, &o \end{pmatrix}_{\text{Gauss.}} = \displaystyle \underset {\begin{psmallmatrix} \mu, & \lambda, &o \end{psmallmatrix}}{\operatorname {arg\;min}  J_{\text{Gauss.}}}
	\end{eqnarray}
	\subsubsection{Case-study}
\begin{figure*}
		\centering
		\fontsize{7.5pt}{12pt}
		\def\svgscale{0.7}
		\includegraphics[width=\textwidth]{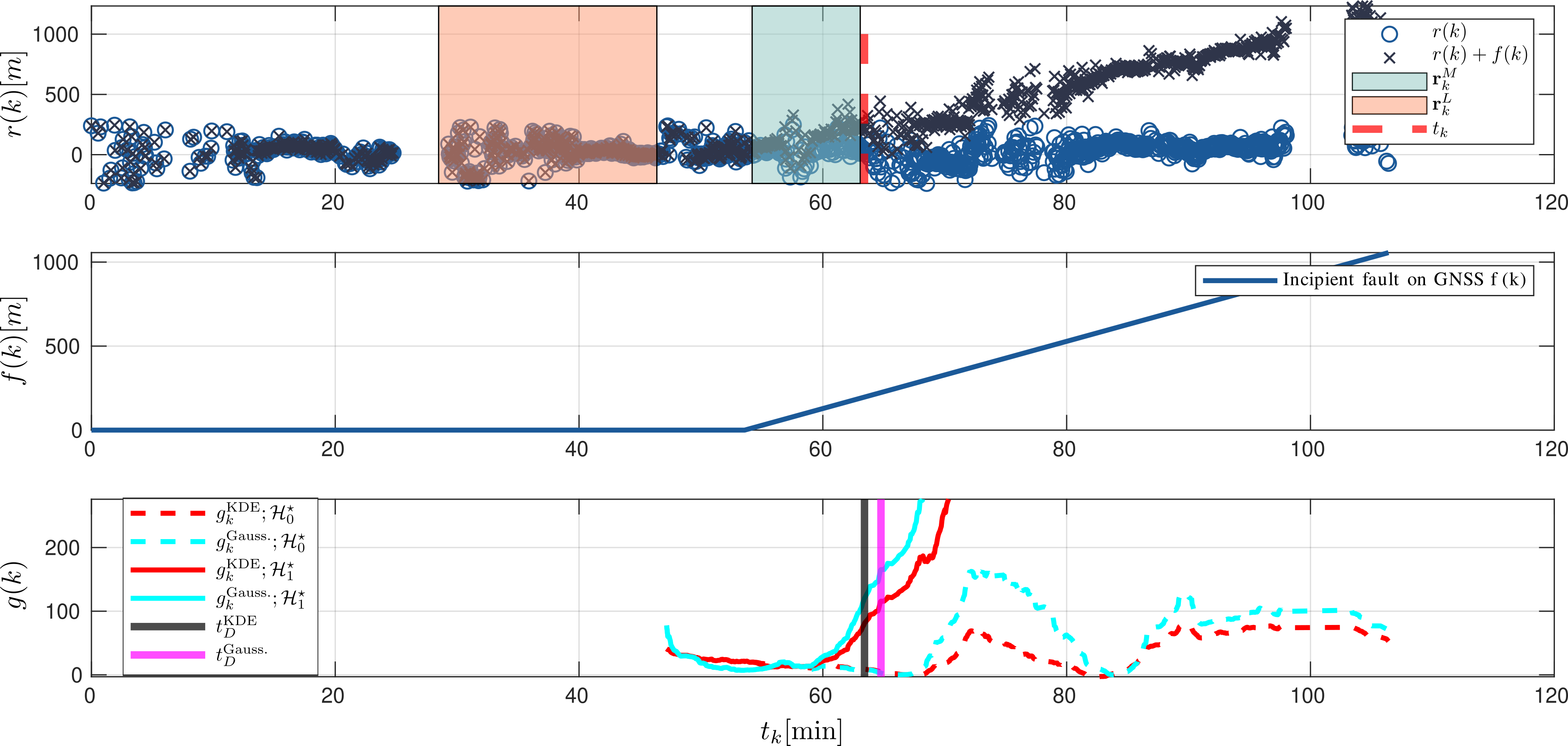}
		\caption{Top: The overall data record $\vec{r}$ and the rolling windows $\vec{r}^L_k,\vec{r}^M_k$. In blue data corresponding to nominal conditions, and in black the same data incorporating a change that is being detected by the GLR in the bottom subfigure. Mid: Spoofing the GNSS measurements with an additive cross-track error $f\ofk=f_{\text{slope}}\cdot (t-t_{\text{onset}})$ . Performance is compared based on the time to detect $t_D$. The detection times for the individual detectors are $t^{\text{KDE}}_D=5.8 \si{min}$, $t^{\text{Gauss.}}_D=9\si{min}$. Bottom: Test statistics $g\ofk$  corresponding to each of the detectors and under both hypothesis. Here, the window lengths are $\mu=9 \si{min},\ \lambda=18 \si{min} $ and the gap between the windows is $o=10 \si{min} $.}
		\label{fig:panel}
	\end{figure*}

\begin{figure}[ptb]
		\centering
		\includegraphics[width=0.8\columnwidth]{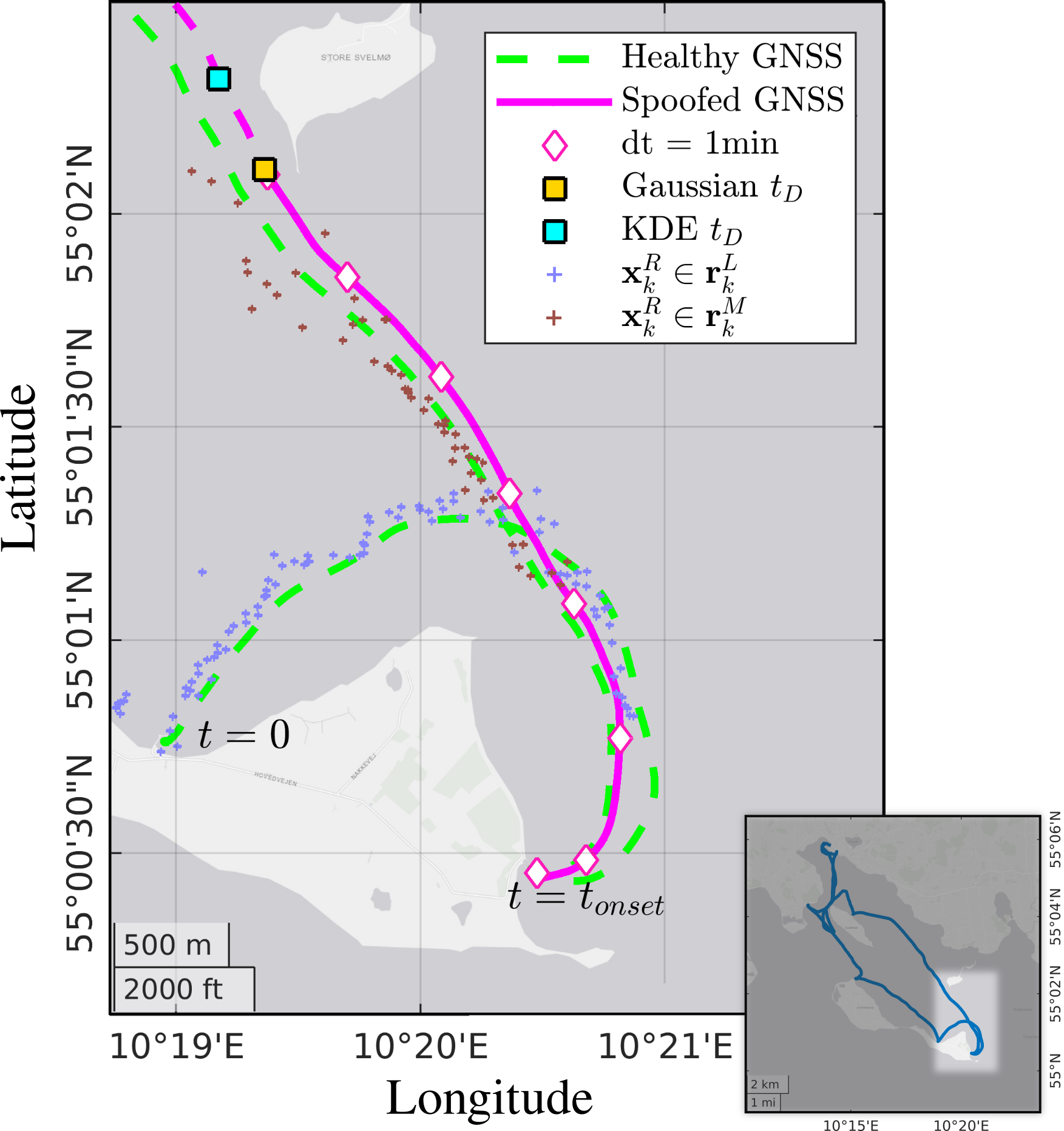}
		\caption{Single realization of a spoofing attack. Comparing performance between the KDE and the Gaussian detector. In the specific illustrated realization, the KDE outperforms the Gaussian detector by a margin of $\SI{4}{\min}$ , or cross-track error $\Delta f= f(t^{\text{ KDE}}_D)- f(t^{\text{Gauss.}}_D)= \SI{120}{\meter}$}
		\label{fig:performance_comparison_geo}
	\end{figure}

	In \Cref{fig:panel,fig:performance_comparison_geo} we chose to highlight a single realization of the different spoofing tests. The difference in performance is  evident in this case, more in specific, concerning the time to detect:
	\begin{enumerate}[label=(\alph*)]
		\item $t^{\text{Gauss.}}_D = 9\si{min}$ for the GLRT using the parametric Gaussian distribution
		\item $t^{\text{KDE}}_D = 5.8\si{min}$ for the GLRT under the non-parametric Kernel Density Estimate distribution.
	\end{enumerate}
	
	In the specific case, the Gaussian assumption about the random process generating the residual is invalidated, thus the KDE detector outperforms the Gaussian one.
	
\begin{figure}[ptb]
		\centering
		\includegraphics[width=\columnwidth]{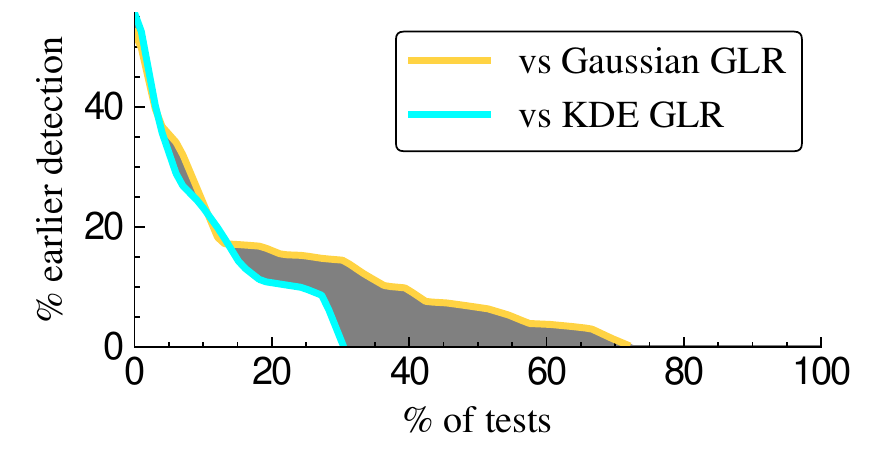}
		\caption{The figure illustrates the margin for performance benefit, of using a combined approach instead of using solely either a Gaussian or a KDE based GLRT.The experiment in  \Cref{fig:performance_comparison_geo} is repeated 1000 times, with the fault's starting time uniformly spread across the missions duration. Figure should read as: " In x\% of tests the combined approach detected the fault at least y\% earlier than the individual detector".}
		\label{fig:statistics_performance}
	\end{figure}

	In an effort to generalize our results, in \Cref{fig:statistics_performance} we present results from all 1000 spoofing-realizations where the performance of the parallel approach is compared to an approach based solely on the KDE or the Gaussian GLRT.
	
	
	\comment{Since the residual corresponds to error in the position of the own ship in meters, then for the given slope of change $f_{\text{ slope}}=20\frac{\text{m}}{min}$, the 5 minute difference in performance between the two detectors, means that the own ship would appear to the navigator as deviating from its true position by a margin of an additional $\delta = 20\times 3.2 \approx 64 \si{m}$, if compared to the detection time of the standard Gaussian GLRT.
		
		One could argue thus, that the non-parametric approach increases the detection capabilities of the detector by 37\% in the specific dataset. The author understands that this is a bold statement, and in reality an extensive cross-validation analysis would have to follow-up in order to confirm the claimed benefits of the method described here.
	}
	\comment{
		\begin{figure}
			\centering
			\includegraphics[width=1\columnwidth]{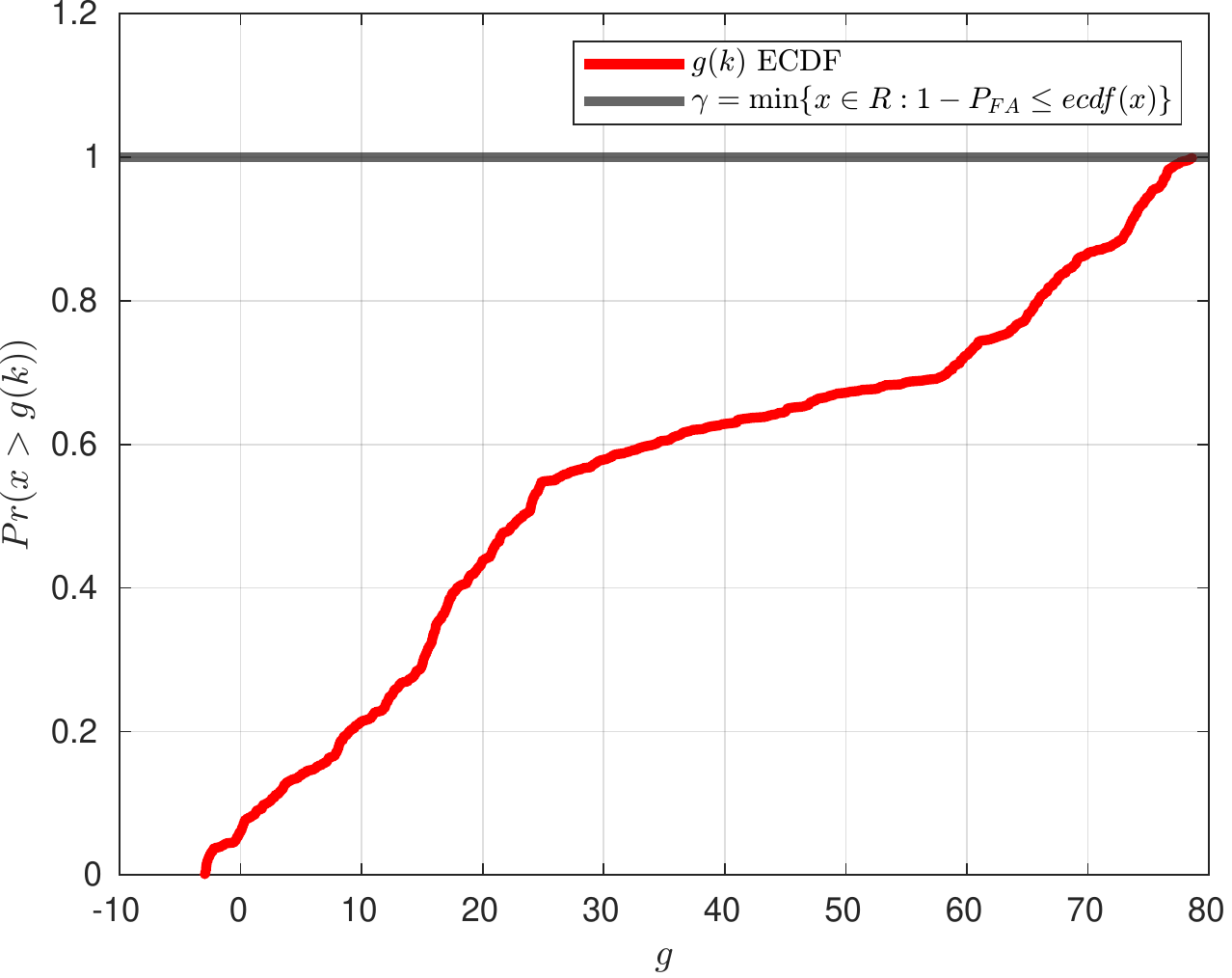}
			\caption{Empirical Cumulative Distribution Function of the test statistic $g\ofk$ under the $\hy_0^\star$ hypothesis and a non-parametric KDE modelling.}
			\label{fig:ecdf}
		\end{figure}
	}

	\subsection{Additional failure modes}
	When it comes to identifying failure modes of a GNSS sensor, we can shortlist them to
	
	\begin{itemize}
		\item Spoofing cyber-attacks, which we identify as the most dangerous
		\item Jamming attacks
		\item GNSS outage due to sensor failure or loss of signal coverage
	\end{itemize}
	
	A jamming attack or outage, will result in no availability of new GNSS measurement, and in the sensor continuously reporting the last available measurement (\cite{Bhatti2017a}). We would like to draw attention on the fact that on such a failure mode, the monitoring system is still able to provide an alarm for the incident. This is a result of the residual $r\ofk$ having a similar profile to the one that we highlighted in our spoofing case study.
	
	Let $t_{k_0}$ be the onset of a spoofing/jamming/outage incident, then the elapsed time from the onset is $\delta t_k = t_k - t_{k_0}$. Moreover, we assume a constant velocity model for a vessel, a more valid assumption the bigger a vessel's displacement is. Given the residual \eqref{eq:residual}
	
	\begin{eqnarray}
	r\ofk =\|\mathcal{T}_{\posegps\ofk} \big(  \posegps\ofk \ -  \pose^R\ofk \big)\|= \omega\big(\posegps\ofk \ -  \pose^R\ofk\big)
	\end{eqnarray}
	For a spoofing incident
	\begin{eqnarray}
	&&\pose^R\ofk \simeq \pose\ofk\\
	&&\posegps\ofk = \pose\ofk + f\ofk\\
	&&r\ofk \simeq\|\mathcal{T}_{\posegps\ofk} \big( f\ofk \big)\|=\omega \big(f_{\text{slope}}\cdot \delta t_k\big)\label{eq:proof1}
	\end{eqnarray}
	For a jamming or outage incident
	\begin{eqnarray}
	&&\pose^R\ofk \simeq \pose\ofk\\
	&&\pose\ofk = \posegps_{k_0} + \dot{\pose}\ofk(t_k-t_{k_0})\\
	&&\posegps\ofk = \posegps_{k_0}\\
	&&r\ofk \simeq \|\mathcal{T}_{\posegps_{k_0}} \big(\dot{\pose}\ofk\delta t_k \big)\| = \omega\big(\dot{\pose}\ofk\delta t_k\big) \label{eq:proof2}
	\end{eqnarray}
	
	From \eqref{eq:proof1},\eqref{eq:proof2} we verify that a jamming or outage incident appears on the residual as a spoofing attack with a slope equal to the vessel's velocity.
	
	\subsection{Conclusion}
	This paper proposed an automated method for absolute positioning in coastal maritime operations, which is independent of GNSS. The method is based on radar scans and the electronic navigation chart. It comprises a two-stage estimation process. The first stage used shoreline-features and a likelihood model, mitigating the need for solving the data association problem. It was argued that the first-stage estimate is always available when land exists within the radar's range. The second stage used detected buoys and beacons at sea and on land, and associated them with similar features on the charts. The output from the first-stage solved the association problem. The availability of the second-stage estimate is limited to areas containing these features. The method was described and its feasibility was demonstrated on real data captured in the South Funen Archipelago (Denmark). The quality of the information delivered by the first-stage estimate was shown to vary, depending on the accuracy of charts and the shoreline temporal variability. This positioning system was shown to provide a redundant system complementary to GNSS. This allowed designing a monitoring system for the GNSS that was used to detect GNSS malfunctions or cyber-attacks, providing immunity to jamming, spoofing or interference. The monitoring systems comprised two parallel Generalized Likelihood Ratio detectors, operating on moving windows. Change detection adopted both parametric and non-parametric distribution modelling. The efficacy of the detector was illustrated by simulated spoofing incidents of the GNSS provided position.
	
	\section{Acknowledgements}
	This research was sponsored by the Danish Innovation Fund, The Danish Maritime Fund, Orients Fund and the Lauritzen Foundation through the Autonomy part of the ShippingLab project, grant number 8090-00063B. Electronic navigational charts were provided by the Danish Geodata Agency.
	%

	\printcredits
	
	\bibliographystyle{cas-model2-names}
	
	\bibliography{bibtex/bib/dd_et_al_journal.bib}
	
	
	
\end{document}